\documentclass[traditabstract]{aa} 

\usepackage{graphicx}
\usepackage{txfonts}
\usepackage{url}
\usepackage{lscape}
\usepackage{longtable}

\begin{document}

   \title{{A Virgo Environmental Survey Tracing Ionised Gas Emission (VESTIGE).III. Star formation in the stripped gas of NGC 4254\thanks{Based on observations obtained with
   MegaPrime/MegaCam, a joint project of CFHT and CEA/DAPNIA, at the Canadian-French-Hawaii Telescope
   (CFHT) which is operated by the National Research Council (NRC) of Canada, the Institut National
   des Sciences de l'Univers of the Centre National de la Recherche Scientifique (CNRS) of France and
   the University of Hawaii.},
   }
   }
   \subtitle{}
  \author{A. Boselli\inst{1}\thanks{Visiting Astronomer at NRC Herzberg Astronomy and Astrophysics, 5071 West Saanich Road, Victoria, BC, V9E 2E7, Canada},         
          M. Fossati\inst{2,3},   
          J.C. Cuillandre\inst{4},
          S. Boissier\inst{1},
	  M. Boquien\inst{5},
	  V. Buat\inst{1},
	  D. Burgarella\inst{1},
          G. Consolandi\inst{6,7},          
          L. Cortese\inst{8},   
          P. C{\^o}t{\'e}\inst{9},
	  S. C{\^o}t{\'e}\inst{9},
          P. Durrell\inst{10},
          L. Ferrarese\inst{9},
          M. Fumagalli\inst{11},  
          G. Gavazzi\inst{6},
          S. Gwyn\inst{9},      
          G. Hensler\inst{12},
	  B. Koribalski\inst{13},
	  J. Roediger\inst{9},
	  Y. Roehlly\inst{14},
	  D. Russeil\inst{1},
          M. Sun\inst{15},
          E. Toloba\inst{16},
	  B. Vollmer\inst{17},
	  A. Zavagno\inst{1}
       }

\institute{     
                Aix Marseille Univ, CNRS, LAM, Laboratoire d'Astrophysique de Marseille, Marseille, France
             \email{alessandro.boselli@lam.fr, samuel.boissier@lam.fr, veronique.buat@lam.fr, denis.burgarella@lam.fr, delphine.russeil@lam.fr, annie.zavagno@lam.fr}
        \and  
                Universit{\"a}ts-Sternwarte M{\"u}nchen, Scheinerstrasse 1, D-81679 M{\"u}nchen, Germany
        \and
                Max-Planck-Institut f\"{u}r Extraterrestrische Physik, Giessenbachstrasse, 85748, Garching, Germany 
                \email{mfossati@mpe.mpg.de}
        \and
                CEA/IRFU/SAP, Laboratoire AIM Paris-Saclay, CNRS/INSU, Université Paris Diderot, Observatoire de Paris, PSL Research University, F-91191 Gif-sur-Yvette Cedex, France
                \email{jc.cuillandre@cea.fr}
	\and
		Universidad de Antofagasta, Unidad de Astronomia, Avenida Angamos 601, Antofagasta 1270300, Chile
		\email{mederic.boquien@uantof.cl}
        \and
                Universit\'a di Milano-Bicocca, piazza della scienza 3, 20100, Milano, Italy
                \email{giuseppe.gavazzi@mib.infn.it}
        \and    
                INAF - Osservatorio Astronomico di Brera, via Brera 28, 20159 Milano, Italy
		\email{guido.consolandi@brera.inaf.it}
        \and
		International Centre for Radio Astronomy Research, The University of Western Australia, 35 Stirling Highway, Crawley WA 6009, Australia
                \email{luca.cortese@uwa.edu.au}
        \and
                NRC Herzberg Astronomy and Astrophysics, 5071 West Saanich Road, Victoria, BC, V9E 2E7, Canada
                \email{laura.ferrarese@nrc-cnrc.gr.ca, patrick.cote@nrc-cnrc.gr.ca, stephen.gwyn@nrc-cnrc.gr.ca, joel.roediger@nrc-cnrc.gr.ca}
        \and
                Department of Physiscs and Astronomy, Youngstown State University, Youngstown, OH, USA
                \email{prdurrell@ysu.edu}
        \and
                Institute for Computational Cosmology and Centre for Extragalactic Astronomy, Department of Physics, Durham University, South Road, Durham DH1 3LE, UK
                \email{michele.fumagalli@durham.ac.uk}
        \and
                Department of Astrophysics, University of Vienna, T\"urkenschanzstrasse 17, 1180, Vienna, Austria
                \email{gerhard.hensler@univie.ac.at}
 	\and
		Australia Telescope National Facility, CSIRO Astronomy and Space Science, P.O. Box 76, Epping, NSW 1710, Australia
		\email{baerbel.koribalski@csiro.au}
	\and
		Astronomy Centre, Department of Physics and Astronomy, University of Sussex, Falmer, Brighton BN1 9QH, UK
        \and
                Physics Department, University of Alabama in Huntsville, Huntsville, AL 35899, USA
                \email{ms0071@uah.edu}
        \and
                Department of Physiscs, University of the Pacific, 3601 Pacific Avenue, Stockton, CA 95211, USA
		\email{etoloba@pacific.edu} 
	\and
		Observatoire Astronomique de Strasbourg, UMR 7750, 11, rue de l'Universit\'e, 67000, Strasbourg, France
		\email{bernd.vollmer@astro.unistra.fr}
                }
               
\authorrunning{Boselli et al.}
\titlerunning{Star formation in the stripped gas of NGC 4254}

   \date{}

 
  \abstract  
{During pilot observations of the Virgo Environmental Survey Tracing Galaxy Evolution (VESTIGE), a blind narrow-band H$\alpha$+[NII] imaging survey of the Virgo
cluster carried out with MegaCam at the CFHT, we have observed the spiral galaxy NGC 4254 (M99). Deep H$\alpha$+[NII] narrow-band and GALEX UV images 
revealed the presence of 60 compact (70-500 pc radius) star forming regions up to $\simeq$ 20 kpc outside the optical disc of the galaxy. These regions are located along 
a tail of HI gas stripped from the disc of the galaxy after a rapid gravitational encounter with another Virgo cluster member that simulations indicate occurred 280-750
Myr ago. We have combined the VESTIGE data with multifrequency data from the UV to the far-infrared to characterise the stellar populations of these regions
and study the star formation process in an extreme environment such as the tails of stripped gas embedded in the hot intracluster medium. The colour, spectral energy
distribution (SED), and linear size consistently indicate that these regions are coeval and have been formed after a single burst of star formation that occurred  
$\lesssim$ 100 Myr ago. These regions might become free floating objects within the cluster potential well, and be the local analogues of compact sources 
produced after the interaction of gas-rich systems that occurred during the early formation of clusters.
 }
   {}
   {}
   {}
   {}
   {}


   \maketitle
%

\section{Introduction}

The environment plays a fundamental role in shaping galaxy evolution. Since the work of Dressler (1980), it
is clear that clusters of galaxies are dominated by early-type systems (ellipticals and lenticulars) which are relatively 
rare in the field (see also Whitmore et al. 1993). It is also well established that late-type galaxies in dense environments
have a lower atomic gas content than their isolated counterparts (Haynes \& Giovanelli 1984; Cayatte et al. 1990; 
Solanes et al. 2001; Gavazzi et al. 2005) with several recent indications that this occurs also for the molecular gas 
phase (Fumagalli et al. 2009; Boselli et al. 2014a) and the dust content (Cortese et al. 2010; 2012a). The gas removal,
which begins in the outer regions and progress inwards, leading to truncated discs (Cayatte et al. 1994; Boselli et al. 2006), 
induces a decrease of the star formation activity, as it is systematically observed in nearby clusters (e.g. Kennicutt 1983; Gavazzi et al. 1998; 
Gomez et al. 2003; Boselli et al. 2014b).

Since the discovery of these systematic differences between galaxies in rich environments from those in the field,
observers, modelers and simulators made great strides in identifying the dominant physical mechanism
responsible for the gas removal and for the subsequent quenching of the star formation activity, as reviewed in 
Boselli \& Gavazzi (2006, 2014). The detailed observations of local galaxies with strong signs of an ongoing
perturbation (Kenney et al. 2004, 2014; Gavazzi et al. 2001; Vollmer et al. 2006, 2008a,b, 2009, 2012; Sun et al. 2007; 
Boselli et al. 2006, 2016a; Scott et al. 2012), the analysis of local samples of cluster galaxies 
(Gavazzi et al. 1998, 2010, 2013; Boselli et al. 2008ab, 2014b, 2016b), as well as finely tuned models and simulations 
of representative objects (Roediger \& Hensler 2005; Roediger \& Bruggen 2007, 2008; Tonnesen \& Bryan 2009) suggest that the dominant process in local clusters is the ram 
pressure (Gunn \& Gott 1972) exerted by the dense intracluster medium on the interstellar medium of galaxies 
moving at high velocity within the cluster. On the other hand, the statistical analysis of large samples of galaxies
extracted from the SDSS spanning a wide range of density, combined with hydrodynamic cosmological simulations,
favour a less violent interaction able to remove only the hot gas halo and thus stop
the the infall of fresh material, thus reducing the activity of star formation only once the the cold gas
on the galaxy discs has been fully transformed into stars (starvation - Larson et al. 1980). 
Other processes, such as galaxy harassment (Moore et al. 1998), might have significantly contributed during the accretion of galaxies in massive clusters 
via small groups (pre-processing, Dressler 2004) and may still be dominant under specific conditions.

So far, however, there has been no systematic efforts to study the fate of the stripped gas. This situation is 
mainly due to two major reasons: 1) the difficulty in observing the gas once removed from the galactic disc, and thus to provide stringent
constraints to models and simulations; 2) the difficulty in simulating the dynamical evolution of the gas properties within a complex and inhomogeneous
cluster environment (density, turbulence, temperature) in its different phases (atomic, molecular, ionised, hot) and at different 
scales (from molecular clouds on pc scales to tails 50-100 kpc long). One of the first detections of stripped material in cluster late-type galaxies 
is the observations of $\sim$ 50 kpc long radio continuum tails (synchrotron emission) in three galaxies at the periphery of A1367 
(Gavazzi \& Jaffe 1985, 1987; Gavazzi et al. 1995). Their cometary shape has been interpreted as a clear evidence of an ongoing ram 
pressure stripping event. Deep narrow-band H$\alpha$ observations of the same galaxies have highlighted long tails of ionised gas (Gavazzi et al. 2001), suggesting that
the cold atomic hydrogen might change phase once in contact with the hot intracluster medium via e.g., thermal evaporation (Cowie \& Songaila 1977). 
Indeed, the observations of tails of neutral atomic gas (HI) are still quite uncommon: only a few cases are known in nearby
clusters (Virgo - Chung et al. 2007; A1367 - Scott et al. 2012). The situation changed recently thanks to the advent of large panoramic detectors
mounted on 4 metre class telescopes with narrow-band filters, that allowed the detection of tails of ionised gas after very deep exposures in several 
galaxies in Virgo (Yoshida et al. 2002; Kenney et al. 2008; Boselli et al. 2016a), Coma (Yagi et al. 2010; Fossati et al. 2012), Norma (Sun et al. 2007; Zhang et al. 2013), 
and A1367 (Gavazzi et al. 2001; Boselli \& Gavazzi 2014; Yagi et al. 2017). A further observational constraint came from X-ray observations, which allowed the detection of hot gas 
within these tails (Sun et al. 2006). It comes also from long-slit (Yoshida et al. 2012; Yagi et al. 2013) or IFU spectroscopic observations 
(Fumagalli et al. 2014; Fossati et al. 2016; Poggianti et al. 2017; Bellhouse et al. 2017; Fritz et al. 2017; Consolandi et al. 2017), which are
fundamental for understanding the kinematic of the stripped gas and its chemical composition and physical state. Recently, also CO observations have been made possible
for the detection of the molecular gas phase (Jachym et al. 2013, 2014, 2017; Verdugo et al. 2015).

A surprising result of these recent studies is that only in a few cases does the stripped gas collapse to form stars outside the disc of the perturbed galaxy.
This generally happens within compact H{\,\sc{ii}} regions dominated by young stellar populations (Hester et al. 2010; 
Fumagalli et al. 2011a; Fumagalli et al. 2014; Fossati et al. 2016; Consolandi et al. 2017).
There are indeed several instances within the Virgo cluster where the stripped gas remains diffuse in its atomic neutral (Boissier et al. 2012) 
or ionised (Boselli et al, 2016a) phase and does not collapse to form new stars, while in other objects a star formation event occurs (Hester et al. 2010;
Fumagalli et al. 2011a; Arrigoni-Battaia et al. 2012; Kenney et al. 2014).  
This observational evidence is in contrast with the results of models and simulations which systematically predict the formation of new stars outside the 
galaxy disc (Kapferer et al. 2009; Tonnesen \& Bryan 2010, 2012), an indication that the physical prescriptions used in the models need still to be refined.

VESTIGE (A Virgo Environmental Survey Tracing Ionised Gas Emission; Boselli et al. 2018, paper I) is a deep narrow-band H$\alpha$ imaging survey of 
the Virgo cluster carried out with MegaCam at the CFHT. This survey is providing us with a unique opportunity 
to study the fate of the stripped gas in cluster galaxies, being the first complete narrow-band imaging survey of a nearby cluster
up to its virial radius (covering an area of 104$^o$$^2$).
Thanks to its unique sensitivity ($f(H\alpha)$ $\sim$ 4 $\times$ 10$^{-17}$ erg sec$^{-1}$ cm$^{-2}$ - 5$\sigma$ detection limit for point sources and
$\Sigma$(H$\alpha$) $\sim$ 2 $\times$ 10$^{-18}$ erg sec$^{-1}$ cm$^{-2}$ arcsec$^{-2}$ - 1$\sigma$ detection limit 
at 3 arcsec resolution for extended sources) and subarcsecond image quality, VESTIGE will detect the diffuse ionised gas
stripped from the perturbed galaxies, as well as compact H{\,\sc{ii}} regions of 
luminosity $L(H\alpha)$ $\geq$ 10$^{36}$ erg s$^{-1}$. VESTIGE will also benefit from the large number of available multifrequency data
for the Virgo cluster, from the X-ray to the radio, which are necessary for a coherent and complete analysis of the multiphase gas (Boselli et al. 2017).  
The use of H$\alpha$ data with respect to other star formation tracers is crucial if we want to study the process of star formation 
on timescales of $\lesssim$ 10 Myr (Boselli et al. 2009; Boquien et al. 2014). This is necessary whenever the perturbing process is 
rapid ($\lesssim$ 500 Myr), as indeed expected within a rich cluster such as Virgo (Boselli et al. 2016b).

This paper is dedicated to the study of the spiral (SA(s)c) galaxy NGC 4254 (M99) located at the northern periphery of the cluster at $\simeq$ 1.0 Mpc 
from M87 ($\simeq$ 0.7 $\times$ $R_{vir}$).
The presence of a $\simeq$ 250 kpc long tail of HI gas indicates that the galaxy has been gravitationally perturbed after a rapid 
encounter with another Virgo cluster member (Haynes et al. 2007). The interaction is thought to be responsible 
for the creation of a free floating HI cloud of gas not associated to any stellar component, the so called dark galaxy Virgo HI21
(Minchin et al. 2005). Dedicated simulations suggest that this interaction occurred $\sim$ 
280 - 750 Myr ago (Vollmer et al. 2005; Bekki et al. 2005; Duc \& Bournaud 2008). The same simulations indicate that, during a rapid galaxy-galaxy 
encounter (galaxy harassment - Moore et al. 1998), only the gaseous component is perturbed leving the stellar disc intact
(Duc \& Bournaud 2008). What happens to the stripped gas? Why is it still in its cold atomic phase? Being still cold, can it collapse and  
form new stars outside the galaxy disc and, if so, under what conditions? NGC 4254 is thus an ideal laboratory 
for studying the star formation process in extreme and unusual conditions, and another candidate to extend previous studies to different
environments (Boquien et al. 2007, 2009, 2010, 2011; Fumagalli et al. 2011a; Arrigoni-Battaia et al. 2012; Jachym et al. 2013, 2014, 2017; 
Kenney et al. 2014; Lisenfeld et al. 2016). 

In this paper, we search for and study 
the properties of extraplanar H{\,\sc{ii}} regions formed after the interaction of the galaxy within the hostile cluster environment. We do that
by using the unique set of multifrequency data
sensitive to the youngest stellar populations available for this representative galaxy combined with SED fitting models. 
The narrow-band H$\alpha$ observations and data reduction are described in Sect. 2.
The multifrequency dataset used in the analysis are described in Sect. 3, while Sect. 4 and 5 describe respectively the 
identification of the extraplanar H{\,\sc{ii}} regions and the derivation of their physical parameters. The discussion is given in Sect. 6.
Given its position within the cluster, we assume the galaxy to be at a distance of 16.5 Mpc (Gavazzi et al. 1999; Mei et al. 2007; Blakeslee et
al. 2009). All magnitudes are given in the AB system.

   \begin{figure*}
   \centering
   \includegraphics[width=19cm]{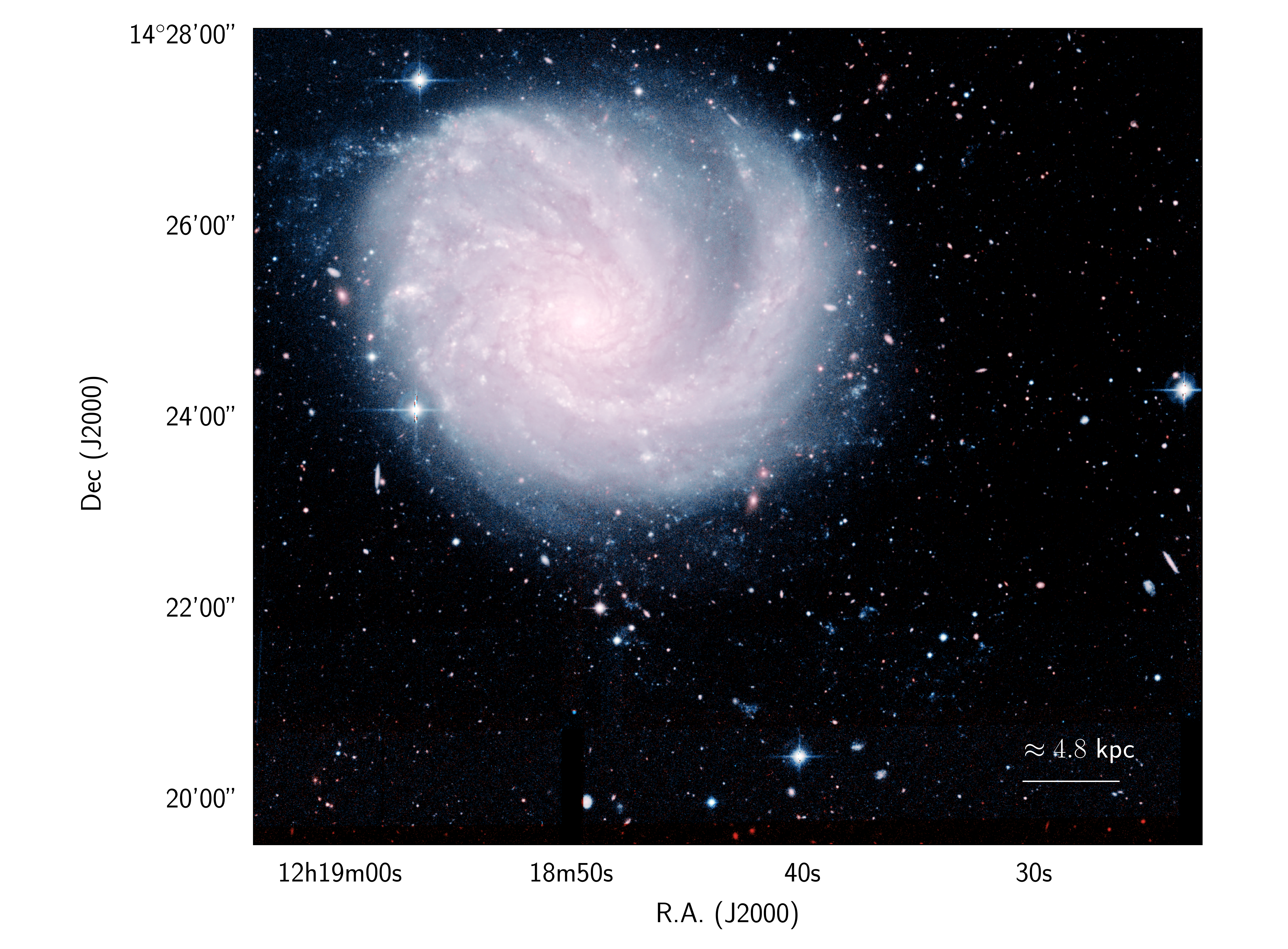}
   \caption{Colour $ugi$ RGB image of the galaxy NGC 4254 obtained using the NGVS data (Ferrarese et al. 2012). At the distance of the galaxy (16.5 Mpc), 1
   arcmin = 4.8 kpc. The star forming regions analysed in this work are the blue blobs in the south-west outside the disc of the galaxy.
 }
   \label{NGVS}%
   \end{figure*}

\section{Observations and data reduction}


The galaxy NGC 4254 has been observed during the second run of pilot observations of the VESTIGE survey in 
spring 2016 (see Boselli et al. 2017 for details). The observations have been carried out with MegaCam at the CFHT using the new narrow-band filter MP9603 ($\lambda_c$ = 6590 \AA; $\Delta \lambda$ = 104 \AA,
$T$ = 93\%) which includes at the redshift of the galaxy ($vel$ = 2404 km s$^{-1}$) the H$\alpha$ line ($\lambda$ = 6563 \AA) and the two [NII] lines
($\lambda$ =6548, 6583 \AA)\footnote{Hereafter we will refer to the H$\alpha$+[NII] band simply as H$\alpha$,
unless otherwise stated.}. The camera is composed of 40 CCDs with a pixel scale of 0.187 arcsec/pixel. 
The galaxy has been observed using the pointing-macro QSO LDP-CCD7 especially designed for the Elixir-LSB data reduction pipeline. This macro is composed of 7 different pointings 
overlapping over a region of 40$\times$30 arcmin$^2$. To secure the determination of the stellar continuum, the galaxy was also observed in the $r$-band filter. 
The integration time for each single pointing within the macro was of 66 sec in the $r$-band and 660 sec in the narrow-band filter. The macro was run two times, thus the total
integration time in the central 40$\times$30 arcmin$^2$ of the combined image is of 924 sec in $r$-band and 9240 sec in narrow-band.



The data reduction has been completed using the standard procedures adopted for the VESTIGE data (extensively described in paper I).
The MegaCam images have been reduced using Elixir-LSB, a data reduction pipeline optimised to detect the diffuse emission of extended low surface brightness features 
associated with the perturbed galaxies by removing any contribution of scattered light in the different science frames. This procedure,
tuned to detect low surface brightness features, 
is perfectly adapted for the narrow-band frames whenever the images are background dominated, as is the case for those obtained in this work (see Boselli et al. 2016a).
The photometric calibration of the $r$-band filter has been derived using the standard MegaCam calibration procedures (Gwyn 2008). Those in the narrow-band filter
have been done as described in Fossati et al. (in prep.). The typical uncertainty is 0.01 mag in both bands. The single images 
were stacked and an astrometric correction was applied using the MegaPipe pipeline (Gwyn 2008). 
As for the VESTIGE survey, the sensitivity is of $f(H\alpha)$ $\sim$ 4$\times$10$^{-17}$ erg s$^{-1}$ cm$^{-2}$ for point souces (5$\sigma$) and 
$\Sigma(H\alpha)$ $\sim$ 2$\times$10$^{-18}$ erg s$^{-1}$ cm$^{-2}$ arcsec$^{-2}$ for extended sources (1$\sigma$ detection limit 
at $\sim$ 3 arcsec resolution), while the depth in the $r$-band is 
24.5 AB mag (5$\sigma$) for point sources and 25.8 AB mag arcsec$^{-2}$ (1$\sigma$) for extended sources. The mean image quality is of 0.60 arcsec in the $r$-band
and 0.56 arcsec in the narrow-band filter.

\section{Multifrequency data}

The Virgo cluster region has been the target of several multifrequency blind surveys. These data are crucial for the identification 
of the stellar population dominating any extraplanar star forming region.

\subsection{GALEX}

Two very deep GALEX exposures are available for the galaxy NGC 4254 in the FUV ($\lambda_c$ = 1539 \AA, $\Delta \lambda$ = 442 \AA; integration time 18131 sec) 
and NUV ($\lambda_c$ = 2316 \AA. $\Delta \lambda$ = 1060 \AA; integration time 27726 sec) bands (Boselli et al. 2011).
For these exposure times, the limiting magnitude for point sources is $FUV$ $\sim$ 25.3 AB mag and $NUV$ $\sim$ 24.9 AB mag (Voyer et al. 2014), while for 
extended sources is $\sim$ 29.7 mag arcsec$^{-2}$ in both bands. The GALEX images have a pixel size of 1.5 arcsec, and an angular resolution of
4.0 arcsec FWHM in the FUV and 5.6 arcsec in the NUV. 
The photometric accuracy
of the instrument is of 0.05 AB mag in FUV and 0.03 in NUV, while the astrometric accuracy is of 0.5 arcsec in both bands (Morrissey et al. 2007).

\subsection{Optical}

Very deep optical images in the $u$, $g$, $i$ and $z$ bands are available from the NGVS survey taken with MegaCam at the CFHT (Ferrarese et al. 2012). 
Their depth is $\sim$ 1 mag deeper than the VESTIGE $r$-band images. The photometric and astrometric accuracy, and angular resolution of NGVS in these photometric bands is 
comparable to that reached in the $r$- and in the narrow-band filters. 

\subsection{Near- and far-IR}

NGC 4254 has been observed in the near- and far-infrared by $\textit{Spitzer}$ with IRAC at 3.6, 4.5, 5.8, and 8.0 $\mu$m 
(Cielsa et al. 2014) and MIPS at 24 $\mu$m and 70 $\mu$m (Bendo et al. 2012) during the SINGS survey (Kennicutt et al. 2003), and
by $\textit{Herschel}$ with PACS at 100 $\mu$m and 160 $\mu$m (Cortese et al. 2014) and SPIRE at 250, 350, and 500 $\mu$m (Ciesla et al. 2012)
as part of the HeViCS (Davies et al. 2010) and HRS (Boselli et al. 2010) surveys. Near-IR data are also available from the WISE survey
(Wright et al. 2010). Given the compact nature of the extraplanar H{\,\sc{ii}} regions analysed in this work, and their weak emission,
the following analysis will use only the four IRAC and MIPS 24 $\mu$m bands, where the angular resolution and the sensitivity of the 
instruments allow the detection or the determination of stringent upper limits for the H{\,\sc{ii}} regions observed in the UV and optical bands. 
None of the sources has been detected at longer wavelengths.      


\subsection{HI and CO}

The discovery of an ongoing perturbation on the disc of NGC 4254 due to a flyby encounter with another Virgo cluster member 
has been made possible by HI observations. VLA observations first revealed the presence of HI blobs located along a diffuse
feature starting at the south-west edge of the stellar disc and extending $\sim$ 30 kpc to the north-west (Phookun et al. 1993). The galaxy has been observed in HI also at Effelsberg
(Vollmer et al. 2005), Parkes (Wong et al. 2006), and with the VLA (Chung et al. 2009).
More recent low angular resolution HI observations taken with the Arecibo radio telescope
during the ALFALFA (Giovanelli et al. 2005) survey of the Virgo cluster indicate that this tail of HI gas is much more extended than previously thought
($\sim$ 250 kpc, to be compared to the isophotal radius of the galaxy which is 13 kpc) and links NGC 4254 to 
the tidal debris Virgo HI21 (Davies et al. 2004; Minchin et al. 2005; Bekki et al. 2005). The total mass of HI gas in this tail
is estimated at (4.8$\pm$0.6) $\times$ 10$^8$ M$_{\odot}$ and its column density $\Sigma (HI)$ $\sim$ 3 $\times$ 10$^{18}$ cm$^{-2}$ (Haynes et al. 2007),
while the HI blobs in the tail detected at the VLA have $\Sigma (HI)$ $\sim$ 10$^{19}$ cm $^{-2}$ at a resolution of $\simeq$ 30 arcsec. The asymmetric distribution of the gas
over the disc of the galaxy, with the presence of a low surface brightness tail in the northern part (Phookun et al. 1993), and the presence
of a radio continuum extended tail (Kantharia et al. 2008), suggest that the galaxy is also suffering ram pressure stripping.\\

NGC 4254 has been also observed in $^{12}$CO(2-1) during the HERACLES survey of nearby galaxies using the IRAM 30m radio telescope, with an angular resolution of 
13 arcsec, a velocity resolution of 2.6 km s$^{-1}$, and a typical RMS of 20-25 mK ($T_{MB}$ scale; Leroy et al. 2009). Unfortunately 
the final map covers only $\sim$ 1/3 of the extraplanar H{\,\sc{ii}} regions of NGC 4254, and does not detect any of them.

\section{Extraplanar H{\,\sc{ii}} regions}

\begin{figure*}
   \centering
   \includegraphics[width=18cm] {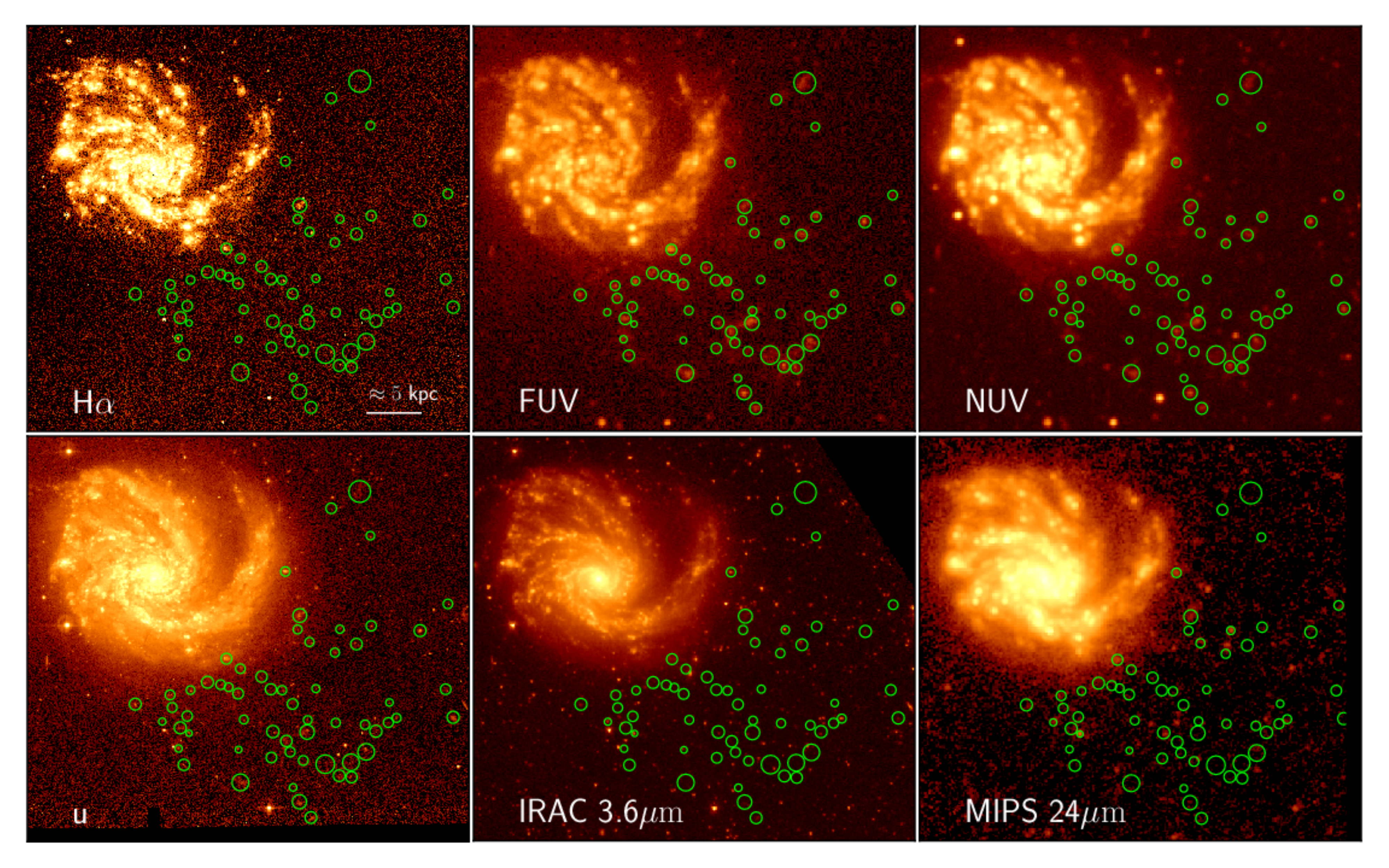}
    \caption{Multifrequency images of the galaxy NGC 4254 (north is up, east is left). Upper panels, from left to right: continuum-subtracted H$\alpha$ (VESTIGE), $FUV$ (GALEX), $NUV$ (GALEX); lower
    panels: $u$-band (NGVS), IRAC 3.6 $\mu$m ($\textit{Spitzer}$), MIPS 24 $\mu$m ($\textit{Spitzer}$).
    The extraplanar star forming regions, marked with green circles, are evident in FUV image at the south-west of the galaxy. The linear size of each single image corresponds 40$\times$35 kpc$^2$ at the distance
    of the galaxy.
 }
   \label{multifrequency}%
   \end{figure*}

\begin{figure*}
   \centering
   \includegraphics[width=18cm] {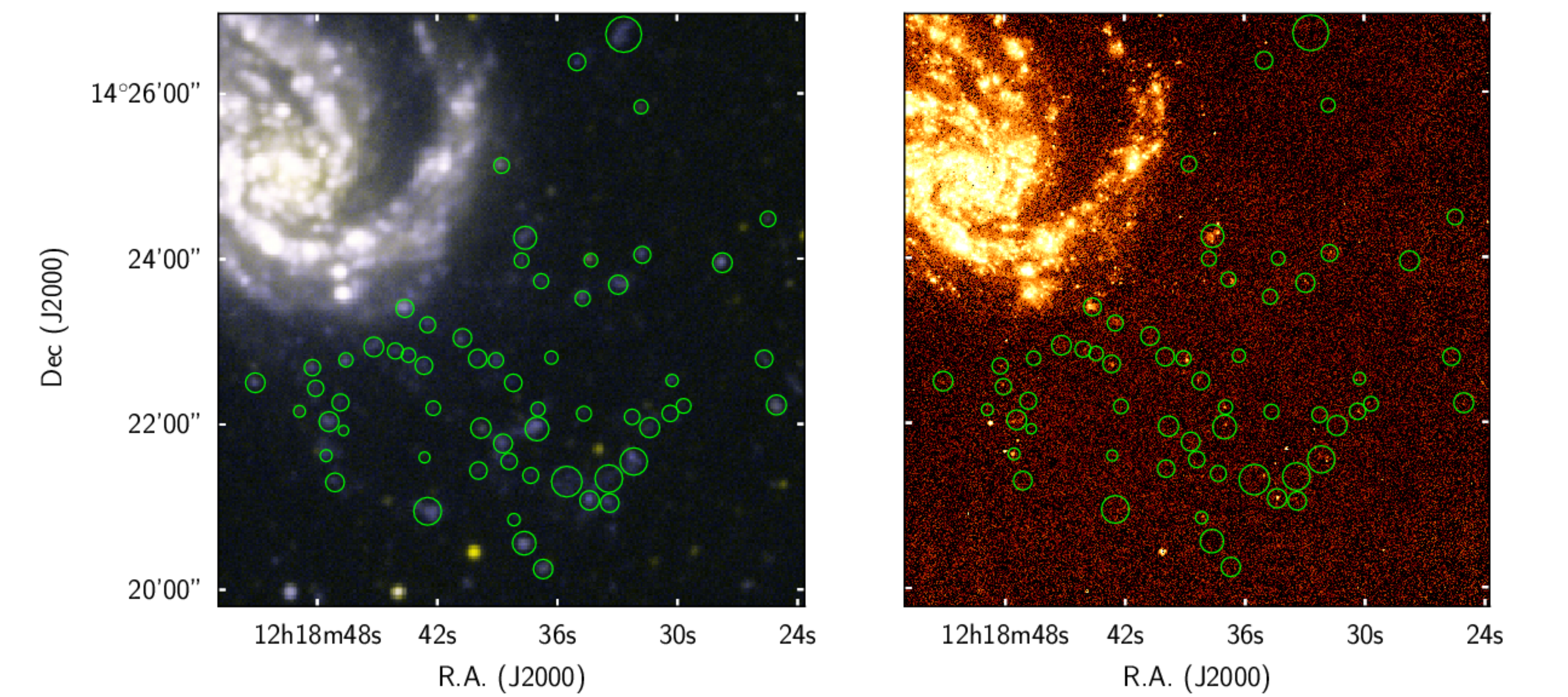}
    \caption{FUV/NUV GALEX colour (left) and continuum-subtracted H$\alpha$ (VESTIGE; right) magnified images of the extraplanar star forming regions (marked with green circles)
    in the south-west quadrant of NGC 4254. The UV colour image shows that these regions have very blue colours, and are thus dominated by young stellar populations ($\lesssim$ 100 Myr). 
    Half of them, however, are undetected in H$\alpha$, indicating that their typical age is 10 Myr $\lesssim$ $age$ $\lesssim$ 100 Myr.
 }
   \label{FUV_Ha_regions}%
   \end{figure*}

Figure \ref{multifrequency} shows the multifrequency images of NGC 4254. The FUV and NUV images indicate the presence of several blue compact 
objects in the south-west periphery of the galaxy extending up to $\sim$ 20 kpc outside the optical disc. These
star forming complexes have been already noticed by Thilker et al. (2007) who identified them as 
part of a standard extended UV disc. Among these we selected in the FUV GALEX image
those located at the edge of the prominent south-west spiral arm or outside the stellar disc of NGC 4254, with a compact morphology, and avoiding extended diffuse structures.
We then rejected those with typical morphologies of background galaxies as seen in the higher quality optical images.  
We identify 60 of these compact regions as depicted in Fig. \ref{FUV_Ha_regions}. These objects are all detected in the NUV image,
30 of them also in the VESTIGE H$\alpha$ continuum-subtracted image, while 34, 26, 24, 21, and 16 in the $ugriz$ images, respectively. 
The VESTIGE and NGVS images indicate that these sources are very compact since only barely resolved 
in these subarcsecond resolution frames. Although this sample is not complete in any sense, we are confident that it is statistically representative 
of the extraplanar young stellar regions associated to NGC 4254.

\begin{figure*}
   \centering
   \includegraphics[width=18cm] {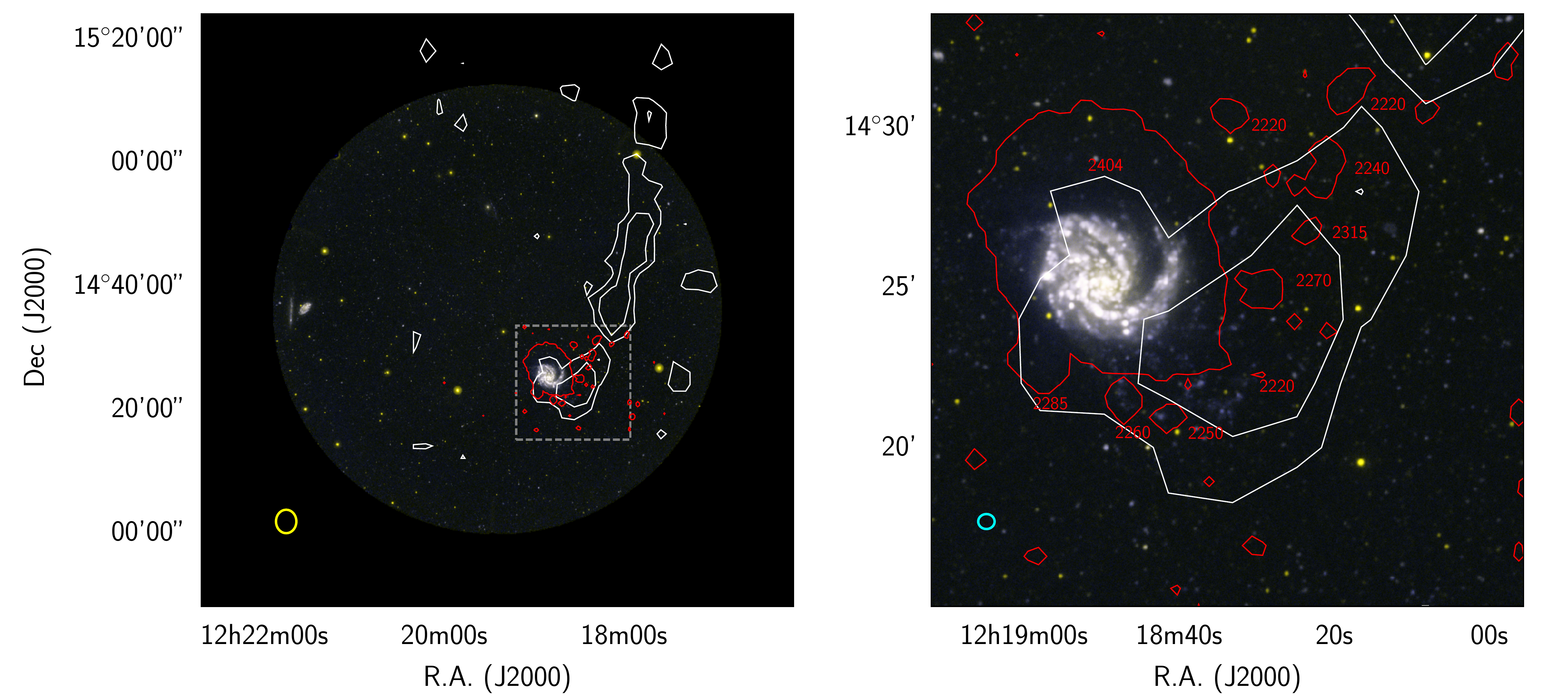}
   \caption{The VLA HI contours at a column density $\Sigma (HI)$ = 10$^{19}$ cm$^{-2}$ (from Chung et al. 2009; red) and the Arecibo
   HI contours at a column density $\Sigma (HI)$ = 1-2 $\times$ 10$^{19}$ cm$^{-2}$ (from Haynes et al. 2007; white) are superposed on the GALEX $FUV/NUV$
   colour image. The right panel is a magnified view of the boxed region marked in the left panel with a dashed line. The typical velocity of the different HI structures detected at
   the VLA are given. The yellow and cyan ellipses at the 
   lower left corner of the two images show the beam sizes at Arecibo (3.3$\arcmin$ $\times$ 3.8$\arcmin$) and at the VLA (30.86$\arcsec$ $\times$ 28.07$\arcsec$).}
   \label{HI}%
   \end{figure*}

The FUV GALEX image alone does not allow us to firmly state whether these blue compact regions are all associated to the galaxy or are foreground or background objects.
However, their relative distribution with respect to NGC 4254 and to the HI tail of stripped gas detected either in the VLA (Chung et al. 2009) or in the ALFALFA HI data
(Haynes et al. 2007), as shown in Fig. \ref{HI}, strongly suggests that they are star forming regions formed within the stripped gas of the galaxy. Surprising 
is the fact that these star forming regions are only present within $\lesssim$ 20 kpc from the edge of the stellar disc of the galaxy, while they are totally absent 
in the rest of the HI tail.
We can estimate the contamination of background objects using the FUV GALEX number counts derived by Xu et al. (2005). At the limiting magnitude of our detections ($FUV$ $\simeq$ 24 mag)
the number of background sources is $\sim$ 1000 deg$^{-2}$, thus $\sim$ 6-7 over the $\sim$ 6' $\times$ 4' region outside NGC 4254 analysed in this work ($\sim$ 10\%).
We also measured the mean background source density down to a limiting magnitude of $FUV$ $\leq$ 24 mag and detected with a signal-to-noise of $SN$ $>$5
within the whole GALEX FUV frame (0.5 deg radius) once the bright Virgo galaxies (NGC 4254, NGC 4262, NGC 4298) have been masked. 
This density is $\sim$ 2300 deg$^{-2}$, thus the expected contamination is $\sim$ 15 objects in the studied region. Since the background distribution slightly 
changes within the FUV frame, we also measured the density of FUV emitting sources with properties similar to those analysed in this work in a corona centered on NGC 4254 
of inner radius 210 arcsec and outer radius 400 arcsec, where all the studied regions are located. Considering as reference the density within this corona, 
the number of expected contaminants is $\sim$ 10 objects.

The coordinate of these regions are given in Table \ref{regiondata}.

To study the stellar populations of these compact regions we extract their fluxes in the different photometric bands using the apertures shown in Fig. \ref{FUV_Ha_regions}.
All these apertures have a radius $r_{ap}$ $\gtrsim$ 4 arcsec to encompass the total flux of the emitting region in the different photometric bands, including
those diffraction limited (FUV, NUV, 24 $\mu$m $\textit{Spitzer}$/MIPS; see Table \ref{regiondata}\footnote{An aperture correction has been applied to the only
region detected at 24 $\mu$m.}). At the same time they are sufficiently small ($r_{ap}$ $\leq$ 10 arcsec, where $r_{ap}$ is the radius of the regions plotted
in Figs. \ref{multifrequency} and \ref{FUV_Ha_regions} encompassing all the $FUV$ emission in the GALEX frame) to limit 
the contamination of background or local point sources. We notice, however, that despite the small size of the aperture, in some cases 
the high resolution MegaCam images in H$\alpha$ or in the broad bands reveal the presence of more than a single source unresolved by GALEX.
None of these regions is detected at wavelengths longer than 24 $\mu$m. To have the most accurate estimate of the flux of each emitting region, or measure a stringent upper limit, the sky background
is derived in the same apertures randomly displaced around the region itself once the other sources, 
and the diffuse emission of the galaxy is masked. This procedure is repeated 1000 times, minimising any
uncertainty on the sky background determination. As in Fossati et al. (2018), the uncertainties on the fluxes are obtained as the quadratic sum of 
the uncertainties on the flux counts and the uncertainties on the background (rms of the bootstrap iterations). The uncertainties on the flux counts  
are derived assuming a Poissonian distribution for the source photons.
We assume as detected those sources with $S/N$ $>$ 3. For the undetected sources, we estimate an upper limit to the flux.
Flux densities, corrected for Galactic extinction following Schlafly \& Finkbeiner (2011) ($E(B-V)$ = 0.034) combined with the reddening law of Fitzpatrick (1999), 
with their uncertainty and 1$\sigma$ upper limits, are listed in Table \ref{regiondata}.  For consistency with the other photometric bands, the H$\alpha$ flux is given as
$LyC$ (in units of $\mu$Jy), the flux in the Lyman continuum pseudo filter $PSEUDO_{LyC}$ derived as described in Boselli et al. (2016b):

\begin{equation}
{LyC ~\rm{[mJy]} = \frac{1.07 \times 10^{-37} \times \it{L(H\alpha)} ~\rm{[erg ~s^{-1}]}}{ \it{D}^2 ~\rm{[Mpc]}}} 
\end{equation}

\section{Physical parameters}

\subsection{H$\alpha$ luminosities}

The narrow-band filter encompasses the H$\alpha$ line at $\lambda$ = 6563 \AA ~ and the two [NII] lines at $\lambda$ = 6548 and $\lambda$ = 6583 \AA.
To measure the H$\alpha$ luminosity of these extraplanar H{\,\sc{ii}} regions shown in Fig. \ref{FUV_Ha_regions} we have first to remove the [NII] contribution.
Since no spectroscopic data are available, we apply a mean standard correction. The typical [NII]$\lambda$6583/H$\alpha$
ratio in extragalactic H{\,\sc{ii}} region is [NII]$\lambda$6583/H$\alpha$ $\lesssim$ 0.3 (e.g. McCall et al. 1985; Kewley et al. 2001; Sanchez et al. 2015), 
with a ratio decreasing radially from the nucleus to the outer disc (Kennicutt et al. 1989; Ho et al. 1997). H{\,\sc{ii}} regions formed within the stripped gas of some perturbed 
galaxies have been observed in nearby clusters. Spectroscopic observations of these peculiar regions give quite different estimates
of the [NII]$\lambda$6583/H$\alpha$ line ratio, ranging from 0.1 to 1 in the tails of two galaxies in the Coma cluster (Yoshida et al. 2012), to $\simeq$ 0.4 
in ESO 137-001 (Fossati et al. 2016), and $\simeq$ 0.2 in the perturbed galaxy JO206 recently observed with MUSE by Poggianti et al. (2017).
Given that the extraplanar H{\,\sc{ii}} regions of NGC 4254 have been formed after the collapse of the stripped gas
which has been removed from the outer disc of the galaxy after a rapid encounter with another cluster member, 
we expect a low metallicity environment. Indeed, the mean metallicity of NGC 4254 is 12+log O/H = 8.73 (Hughes et al. 2013), but the galaxy has a steep metallicity gradient 
suggesting that the metallicity in the outer disc is 12+log (O/H) $\sim$ 8.5 (Skillman et al. 1996). This metallicity also corresponds
to that observed in the outer discs of spiral galaxies (12+log (O/H) $\sim$ 8.45, Sanchez-Menguiano et al. 2016; Bresolin 2017).
We thus assume for the correction [NII]$\lambda$6583/H$\alpha$ = 0.2 (Pettini \& Pagel 2004). 
We also assume a mean dust attenuation of $A(H\alpha)$ = 0.7, a value consistent with those observed
in the extraplanar H{\,\sc{ii}} regions of the same perturbed galaxies (0$\lesssim$ $A(H\alpha)$$\lesssim$ 1 mag - Fossati et al. 2016; Poggianti et al. 2017), 
in the outer discs of late-type galaxies (0$\lesssim$ $A(H\alpha)$$\lesssim$ 2 mag - Kennicutt et al. 1989; Sanchez et al. 2015),
or in UV extended discs ($A(H\alpha)$ $\simeq$ 0.7 mag - Bresolin et al. 2009, 2012). 

   \begin{figure}
   \centering
   \includegraphics[width=7cm]{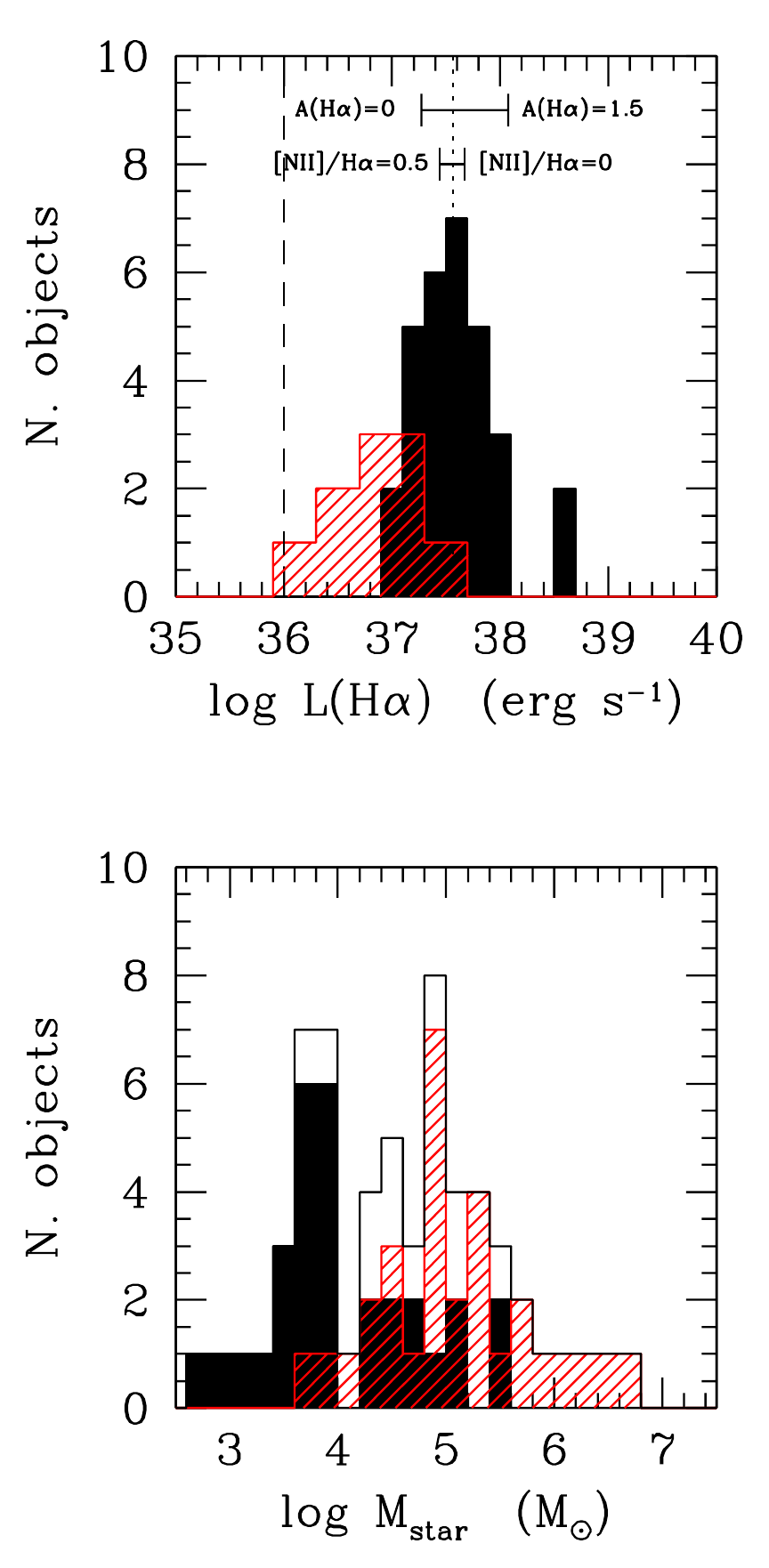}
   \caption{Upper panel: the H$\alpha$ luminosity distribution of the extraplanar H{\,\sc{ii}} regions of NGC 4254. The black shaded histogram 
   indicates the H$\alpha$ detected regions, the red histogram the distribution of the upper limits. The vertical dashed line shows the
   detection limit of VESTIGE, the vertical dotted line the median value for the H$\alpha$ luminosity distribution of the detected sources. 
   The error bars show how the median value of the distribution would shift assuming different corrections for dust attenuation ($A(H\alpha$ = 0.0 -1.5)
   or [NII] contamination ([NII]$\lambda$6583/H$\alpha$ = 0.0 - 0.5). Lower panel: 
   the distribution of the inferred stellar mass. The black open histogram shows the distribution of all the H{\,\sc{ii}} regions. }
   \label{Hadist}%
   \end{figure}

Figure \ref{Hadist} shows the distribution of the H$\alpha$ luminosity of the detected and undetected extraplanar
H{\,\sc{ii}} regions of NGC 4254. Figure \ref{Hadist} also shows how the distribution would 
shift assuming different corrections for dust attenuation ($A(H\alpha)$ = 0.0 - 1.5) or [NII] contamination ([NII]$\lambda$6583/H$\alpha$ = 0.0 - 0.5). 
The luminosity of the  H{\,\sc{ii}} regions is in the range 10$^{37}$ $\lesssim$ $L(H\alpha)$ $\lesssim$ 10$^{39}$ (erg s$^{-1}$),
and corresponds thus to the typical luminosity of giant or super-giant Galactic or extragalactic H{\,\sc{ii}} regions (Lee et al. 2011).




\subsection{Colour analysis}

   \begin{figure}
   \centering
   \includegraphics[width=7cm]{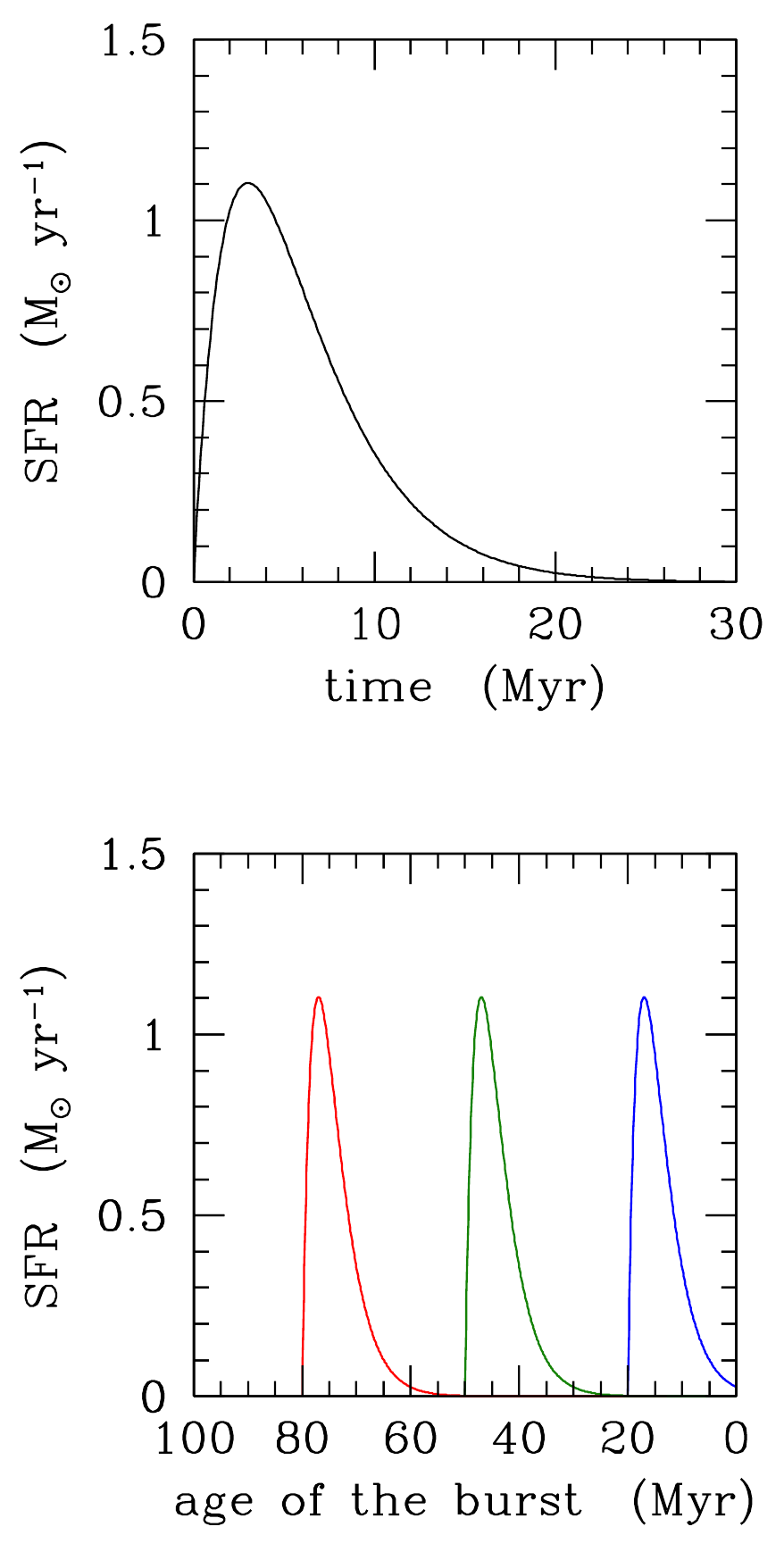}
   \caption{Upper panel: the delayed star formation history with an e-folding time of 3 Myr assumed to fit the observed SED of the extraplanar H{\,\sc{ii}} regions of NGC 4254. 
   Lower panel: the typical age of their dominant stellar population is the age of the burst (80 Myr - red, 50 Myr - green, 20 Myr - blue). }
   \label{delayed}%
   \end{figure}

\begin{figure}
   \centering
   \includegraphics[width=9cm]{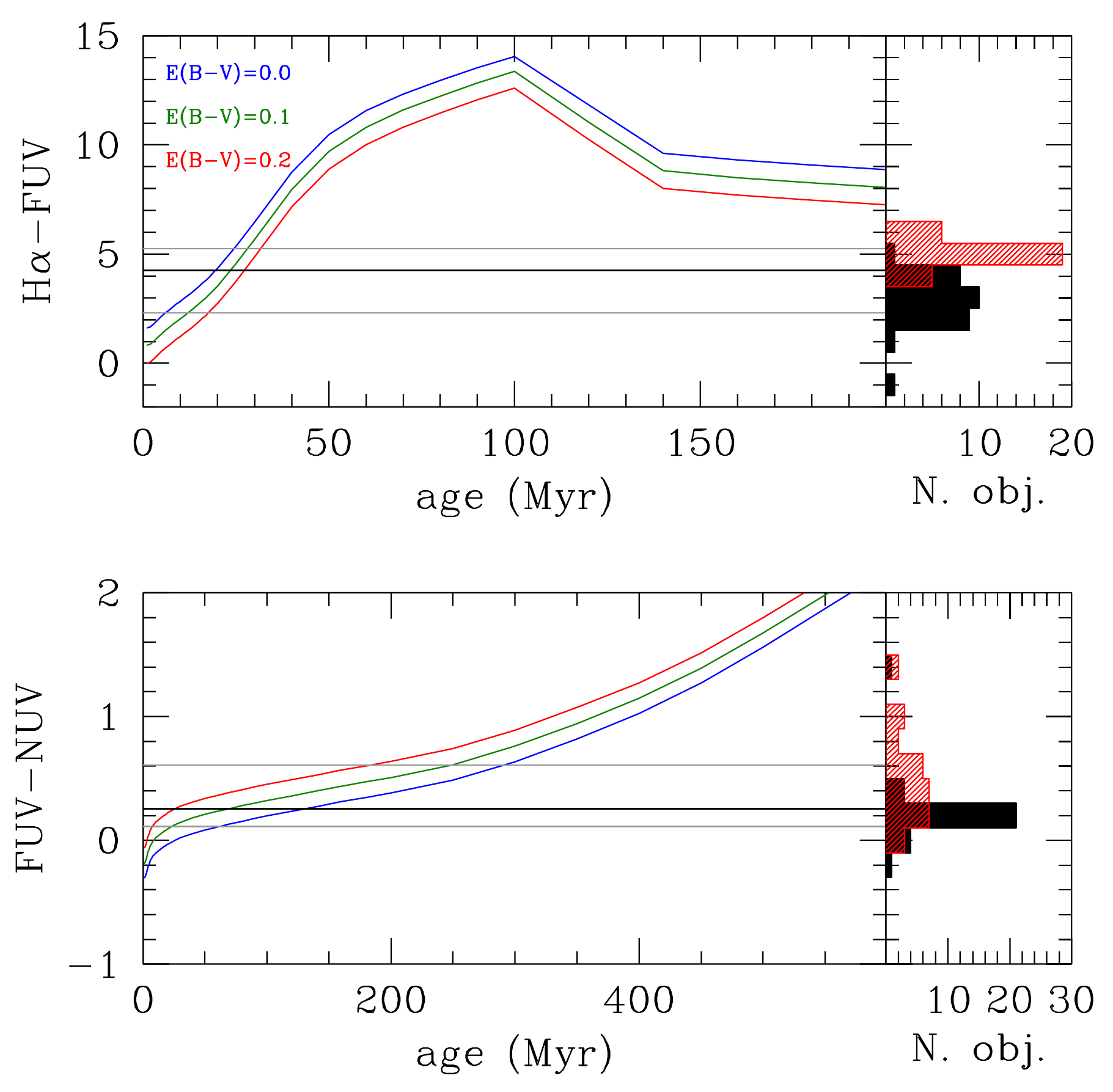}
   \caption{Left panels: variations of the H$\alpha$-$FUV$ (upper) and $FUV-NUV$ (lower) age-sensitive colour indices as a function of time derived for the star formation history given in eq. 1.
   Different colours are used for three different dust attenuations: $E(B-V)_{young}$ = 0.0 (blue), 0.3 (green), and 0.6 (red). The black solid line shows the median of the observed colour distribution, the 
   grey lines the 16\% ~ and the 84\%~ quartiles of the distribution.
   Right panels: distribution of the two colour indices for the detected (black) and undetected (red - lower limit)
   H{\,\sc{ii}} regions. 
   }
   \label{age}%
   \end{figure}

The blue colours of the H{\,\sc{ii}} regions seen in Figures \ref{multifrequency} and  \ref{FUV_Ha_regions} indicate that they are dominated by very young stellar populations. 
The fact that they are all detected in the two UV bands, while only a few are in the H$\alpha$ and $u$-band, and most undetected at longer wavelengths
suggests that they have been formed by a single coeval, short-lived burst of star formation.
To estimate their typical age we first compare their observed ($H\alpha$ - $FUV$) and ($FUV$ - $NUV$) age-sensitive colours 
to the typical synthetic colours of H{\,\sc{ii}} regions derived assuming a star formation history defined as:

\begin{equation}
{SFR(t) = t \times e^{-\frac{t}{\tau}}  ~~~~~\rm{M_{\odot}yr^{-1}}}
\end{equation}

\noindent
the form of which is shown in Fig. \ref{delayed} (upper panel). The $\tau$ e-folding time is set to 3 Myr, consistent with the typical age of giant 
($L(H\alpha)$ $\simeq$ 10$^{37}$ erg s$^{-1}$) H{\,\sc{ii}} regions in the Milky Way
or in other nearby galaxies (Copetti et al. 1985; Tremblin et al. 2014). This approach is similar to the one adopted by Fumagalli et al. (2011a) 
for dating the H{\,\sc{ii}} regions of IC3418 and Boquien et al. (2007) for the intergalactic star forming regions around NGC 5291. 
We then estimate the typical colour of the synthetic H{\,\sc{ii}} regions assuming that the burst of star formation occurred at different epochs
(Fig. \ref{delayed}, lower panel). The synthetic colours are derived using the CIGALE SED fitting code (Noll et al. 2009), assuming the Bruzual \& Charlot (2003)
population synthesis models for the stellar emission with a Salpeter IMF and an updated version of the Draine \& Li (2007) physical models of 
dust emission (see Boselli et al. 2016b for details). To use the H$\alpha$ emission as a constraint in the fit, we convert the extinction corrected H$\alpha$ fluxes into number of
ionising photons as in Boselli et al. (2016b). We assume an escape fraction of the ionising radiation equal to zero, and that the ionising radiation is not absorbed by dust before ionising
the gas (e.g. Boselli et al. 2009). The models are determined assuming different values for the dust attenuation of the stellar continuum ($E(B-V)$ = 
0.0, 0.1, 0.2), as expected if their metallicity and dust content change from region to region. Geometrical effects on the stellar continuum related to the age of the
stellar population are taken into account by using the Calzetti et al. (2000) attenuation law.
To combine in a colour index the H$\alpha$ line flux emission to the broad-band monochromatic flux $FUV$ (in units of mag)
we define the $H\alpha$-$FUV$ colour index as:

\begin{equation}
{H\alpha-FUV = -2.5 \times log(LyC) + 20 - FUV}
\end{equation}

\noindent
where $LyC$ is the flux in the Lyman continuum pseudo filter $PSEUDO_{LyC}$ (in units of $\mu$Jy) as defined in eq. (1). 
Figure \ref{age} shows how the $H\alpha$-$FUV$ and $FUV$-$NUV$ colour indices vary as a function of the age of the burst. 
Clearly the two colour indices are sensitive to the age of the H{\,\sc{ii}} regions, the $H\alpha$-$FUV$ to ages $\lesssim$ 150 Myr, the $FUV$-$NUV$ to ages in the range 500 $\lesssim$ age
$\lesssim$ 1000 Myr. Only the former can be used on the present set of data: the $FUV$-$NUV$ colour index, indeed,
gradually increases from $FUV$-$NUV$ $\simeq$ 0 mag at $age$ = 0 Myr to $FUV$-$NUV$ $\simeq$ 0.6 mag at $age$ = 1000 Myr, a too limited variation compared to the observed distribution of the
$FUV$-$NUV$ colour index and its large uncertainty ($\simeq$ 0.2 mag).  
Figure \ref{age} suggests that the age of the extraplanar H{\,\sc{ii}} regions of NGC 4254 is $\lesssim$ 100 Myr. This is also the case for the H{\,\sc{ii}} regions
undetected in H$\alpha$ ($age$ $\gtrsim$ 10 Myr) and in the $u$-band ($age$ $\lesssim$ 200 Myr).

\begin{figure}
   \centering
   \includegraphics[width=9cm]{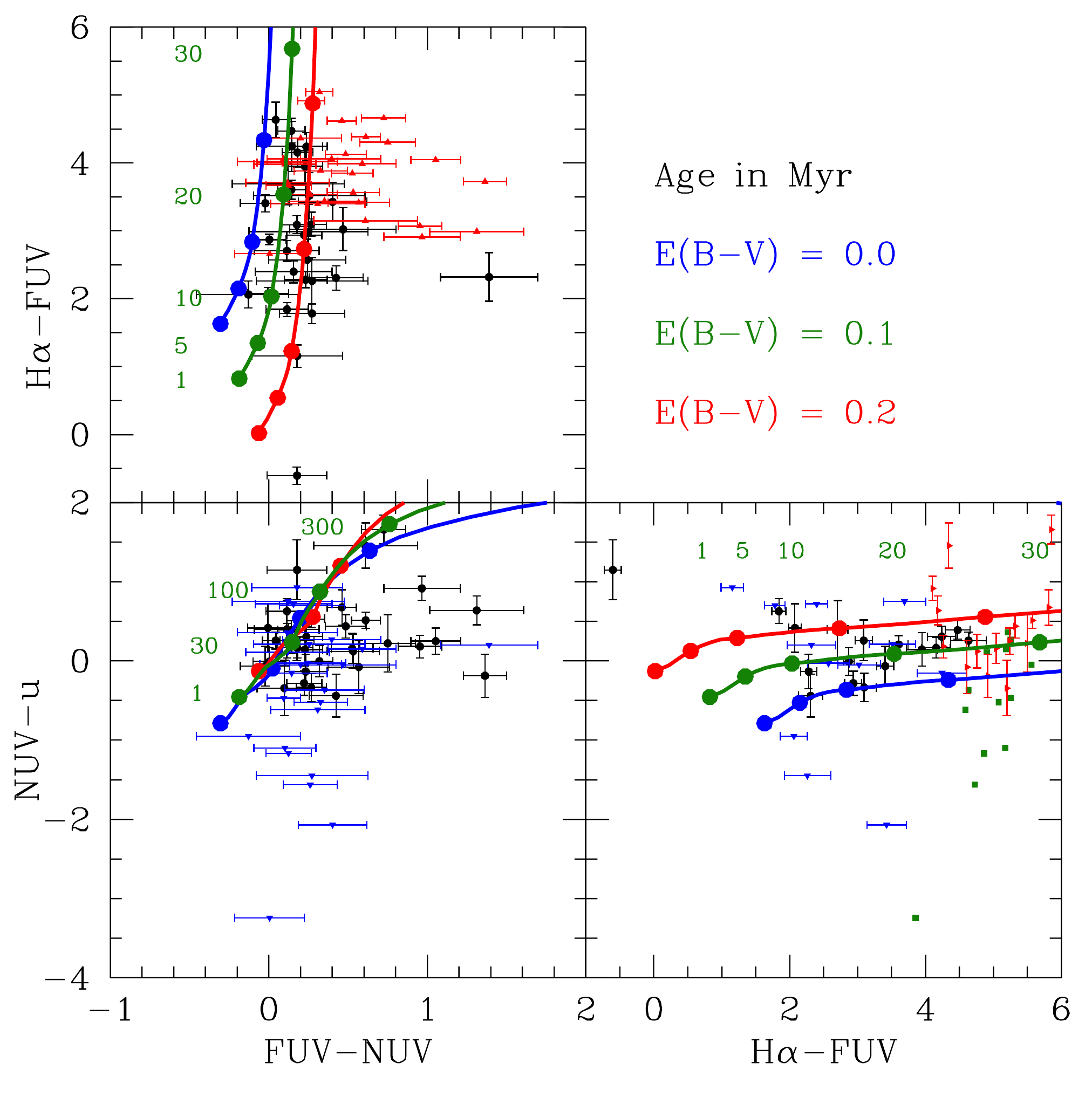}
   \caption{Age-sensitive colour-colour diagrams. Upper left: H$\alpha$-$FUV$ vs. $FUV-NUV$, lower left: $NUV-u$ vs. $FUV-NUV$,
   lower right: $NUV-u$ vs. H$\alpha$-$FUV$. Black filled dots are for regions detected in all bands, red filled triangles
   are lower limits in H$\alpha$-$FUV$, blue filled triangles upper limits in $NUV-u$, and green open squares regions undetected in H$\alpha$ and in the $u$-band. 
   The blue, green, and red filled dots
   and solid lines indicate the expected colours at different ages (in Myr) for the assumed star formation history for an $E(B-V)_Y$ = 0.0, 0.3, and 0.6, 
   respectively.
 }
   \label{FUVNUVHa}%
   \end{figure}

The observed colours of these H{\,\sc{ii}} regions can also be compared to the evolutionary tracks derived by CIGALE in different photometric bands sensitive to young stellar populations,
H$\alpha$, $FUV$, $NUV$, and $u$ (Fig. \ref{FUVNUVHa}). Again, despite the large photometric uncertainties on the data, or the large number of undetected sources in 
some of the photometric bands, the colours of the H{\,\sc{ii}} regions are all consistent with very young ages (0-50 Myr). Figure \ref{FUVNUVHa} also shows the presence of several
outliers with colours significantly different than those predicted by the evolutionary tracks. A clear example are those with $FUV-NUV$ $>$ 0.4 in the upper panel. The visual inspection 
of those detected in H$\alpha$ on the high resolution MegaCam images revealed the presence of multiple sources unresolved in the GALEX frames. If these single H{\,\sc{ii}} regions
have different star formation histories, it is plausible that their UV colour in the unresolved GALEX images does not follow the proposed evolutionary tracks.  

\subsection{SED fitting}

We estimate the age of each single H{\,\sc{ii}} region using the CIGALE SED fitting code assuming the same configuration given above, i.e. the 
star formation history given in eq. (1), the Bruzual \& Charlot (2003) population synthesis models derived for a Salpeter IMF, coupled with the Draine \& Li (2007)
dust models. We fit all (11) photometric bands from the $FUV$ to the \textit{Spitzer}/MIPS 24 $\mu$m with the exception of the $r$-band where the contamination of the
emitting H$\alpha$ line can be dominant combined with the H$\alpha$ emission using 
the $PSEUDO_{LyC}$ pseudo filter as in Boselli et al. (2016b), and treating upper limits as in Sawicki (2012). Upper limits in the far-infrared bands are used to
constrain the dust attenuation in the star forming regions.
The grid of models is created using the parameters listed in Table \ref{cigale} (244620 models).

\begin{figure}
   \centering
   \includegraphics[width=8cm] {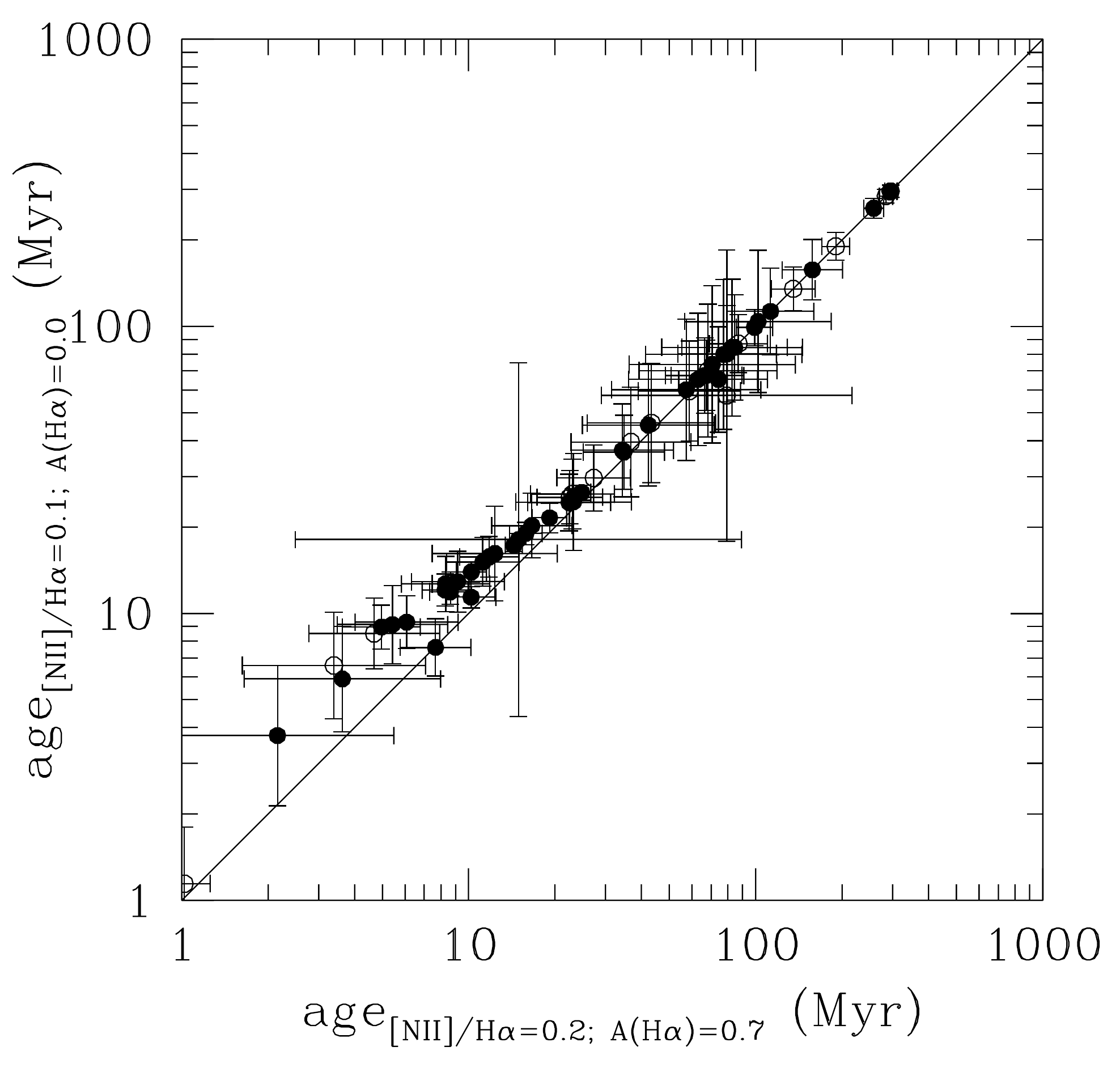}
   \caption{Relationship between the age determined assuming [NII]$\lambda$6583/H$\alpha$ = 0.2 and $A(H\alpha)$ = 0.7 (X-axis) and that assuming 
   [NII]$\lambda$6583/H$\alpha$ = 0.1 and $A(H\alpha)$ = 0.0 (Y-axis). Filled dots indicates the H{\,\sc{ii}} regions where the SED fitting gives reduced $\chi^2_r$ $<$ 6,
   empty dots $\chi^2_r$ $\geq$ 6. The black solid lines show the 1:1 relations.
   }
   \label{agerobustness}%
   \end{figure}

\begin{figure*}
   \centering
   \includegraphics[width=0.32\textwidth] {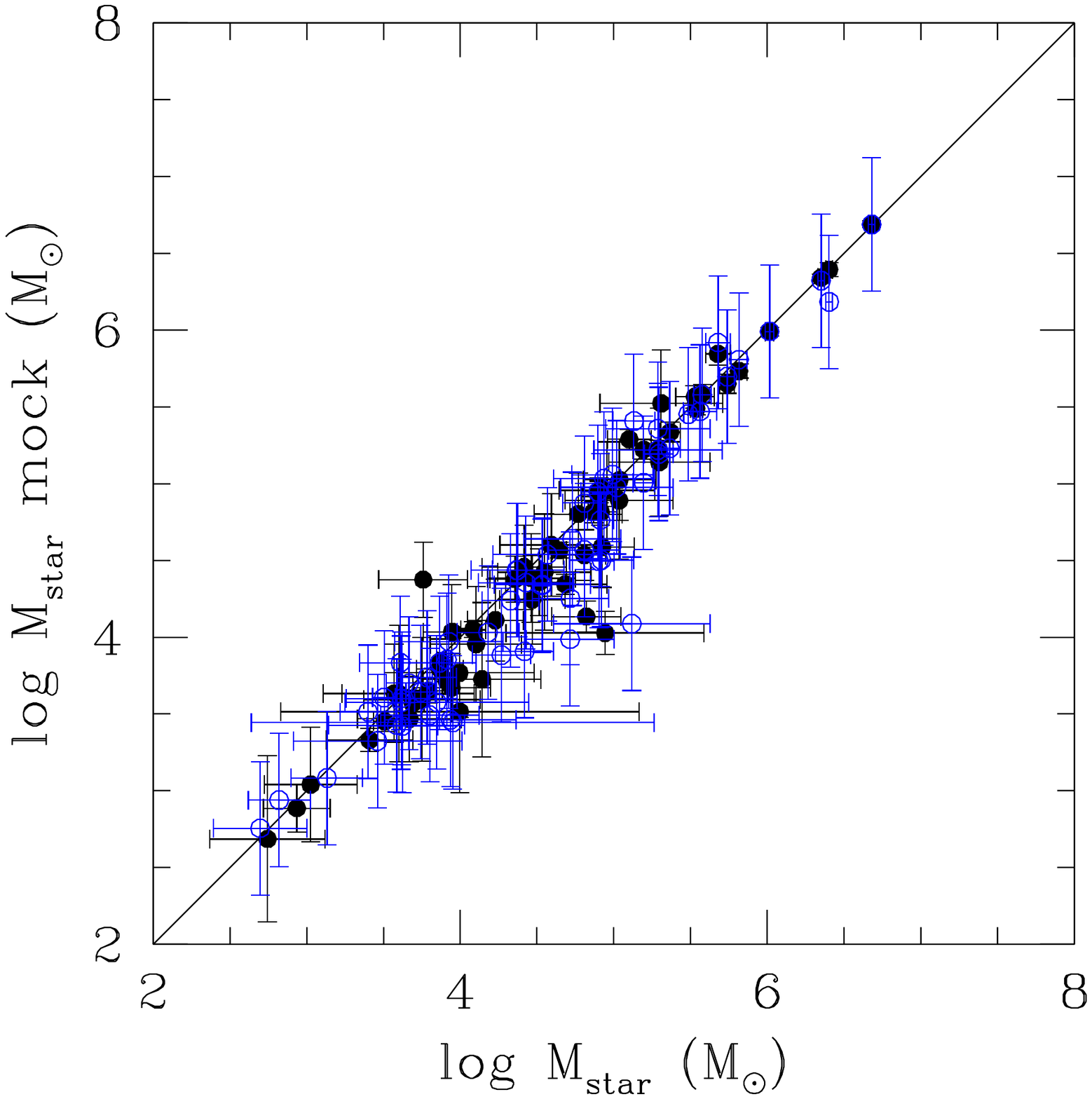}
   \includegraphics[width=0.32\textwidth] {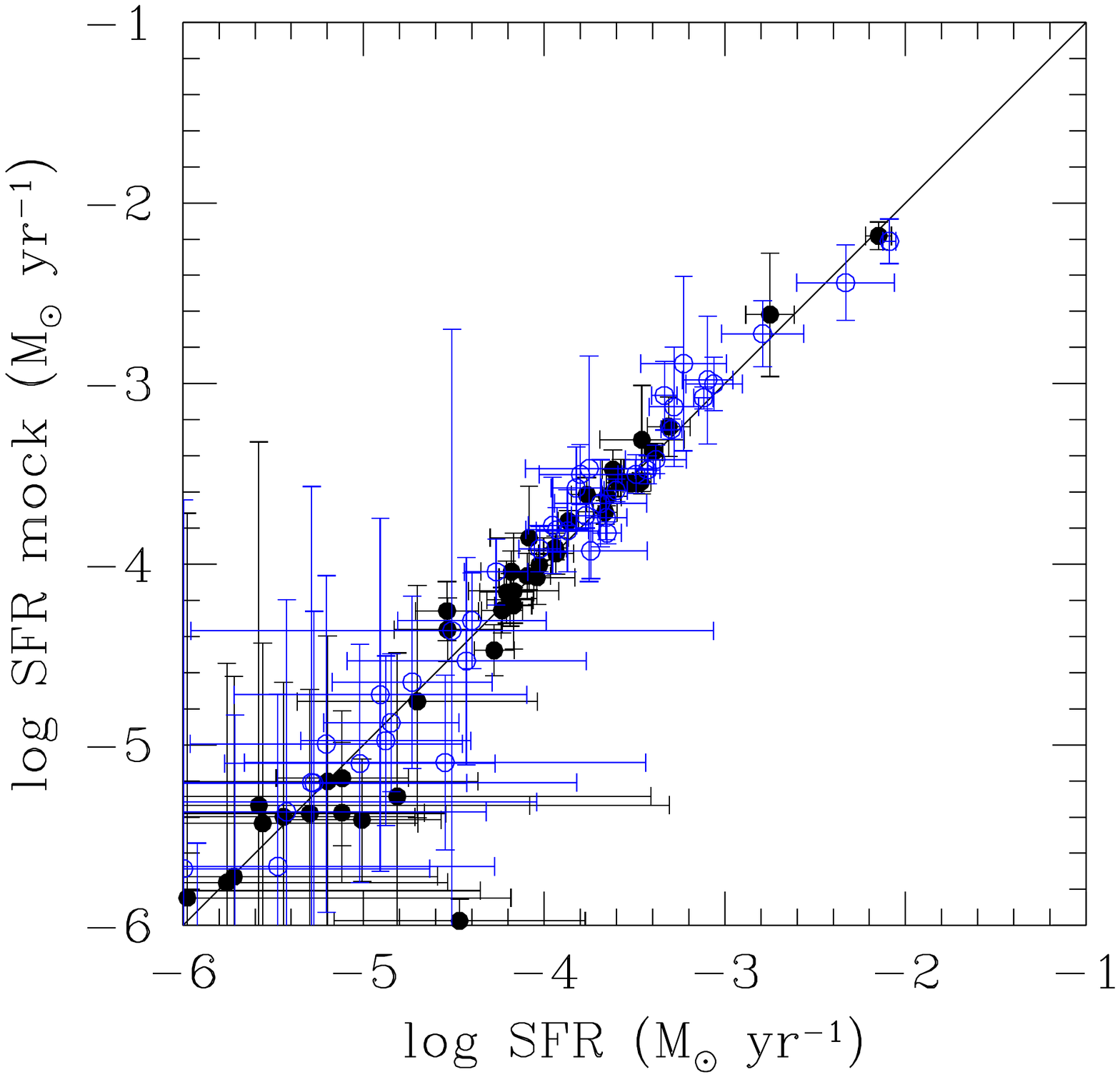} 
   \includegraphics[width=0.32\textwidth] {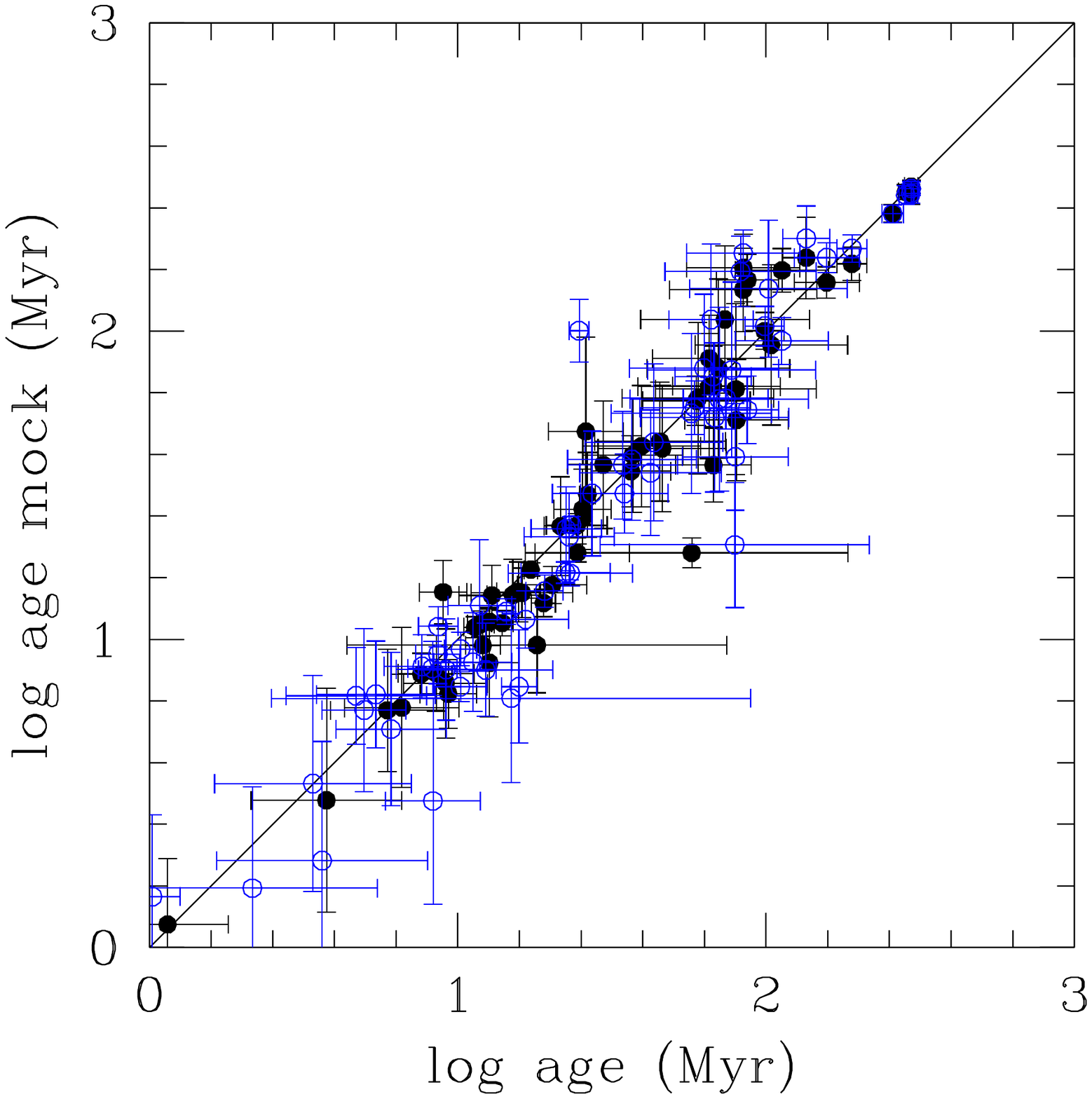}\\
   \caption{Relationship between the stellar mass (left), the star formation rate (centre), and the mean age of the stellar population (right) 
   derived with CIGALE on the observed data (abscissa) and that derived on the mock 
   catalogue as described in Sect. 5.3 (ordinate). Black filled dots are for H$\alpha$ fluxes corrected for [NII] contamination and dust attenuation
   assuming [NII]$\lambda$6583/H$\alpha$ = 0.2 and $A(H\alpha)$ = 0.7 (see Sect. 5.1), blue open circles assuming [NII]$\lambda$6583/H$\alpha$ = 0.1 and $A(H\alpha)$ = 0.0.
   The black solid lines show the 1:1 relations.
   }
   \label{mock}%
   \end{figure*}

The output of the fit (star formation rates, stellar masses, ages, and their uncertainties, self-consistently determined assuming the star formation history
given in eq. 1) for each single H{\,\sc{ii}} region are given in Table
\ref{cigale_out}. 
To quantify how the ages of the regions depend on the assumed dust attenuation and [NII] contamination corrections on the H$\alpha$ flux we compare in Fig.
\ref{agerobustness} two sets of values derived assuming in the first one [NII]$\lambda$6583/H$\alpha$ = 0.1 and $A(H\alpha)$=0.0, in the second [NII]$\lambda$6583/H$\alpha$ = 0.2 and $A(H\alpha)$=0.7.  
We also test the reliability of the output parameters and their uncertainties given 
in Table \ref{cigale_out} by creating a photometric mock catalogue introducing in the observed data some extra noise 
randomly distributed according to a Gaussian curve of standard deviation equal to the median error for each band, and re-fit the data
using the same SED models (see Giovannoli et al. 2011, Boquien et al. 2012, and Ciesla et al. 2014 for details). 
We follow this step twice, the first time keeping the quite uncertain [NII] contamination and H$\alpha$ attenuation 
given in Sect. 5.1 ([NII]$\lambda$6583/H$\alpha$ = 0.2, $A(H\alpha)$ = 0.7), and the second time assuming more extreme values ([NII]$\lambda$6583/H$\alpha$ = 0.1, 
$A(H\alpha)$ = 0.0), and compare the output of the fit in Fig. \ref{mock}. 
Figures \ref{agerobustness} and  \ref{mock} clearly show that, despite the number of parameters used to create the model SEDs, the large uncertainty on the
photometric data and that on the corrections on the [NII] contamination and on the H$\alpha$ attenuation, the stellar masses and the ages of the H{\,\sc{ii}} regions 
derived by the SED fitting code do not vary significantly. 
The mean age of the stellar population is systematically underestimated whenever $A(H\alpha)$ is overestimated. the opposite is true when [NII]/H$\alpha$ is overestimated.  
Star formation rates are instead very uncertain once they are below 10$^{-4}$ M$_{\odot}$
yr$^{-1}$. The SED fitting analysis thus confirms the results obtained from the colour analysis done in the previous section, 
i.e. that these H{\,\sc{ii}} regions have been formed with a few exception $\lesssim$ 100 Myr ago. Some of these exception might be background galaxies.

\subsection{Dynamical age}

It has been shown that the size of the H{\,\sc{ii}} regions is tightly related to their dynamical age, defined as $t_{dyn}$ $\sim$ $r_{HII}$/$\Delta V$, where $r_{HII}$
is the radius of the H{\,\sc{ii}} region and $\Delta V$ its expansion velocity (Ambrocio-Cruz et al. 2016). 
We do not have any dynamical data for these H{\,\sc{ii}} regions, but thanks
to the excellent quality of the H$\alpha$ image we can measure their angular size and compare their stellar age-size relation to 
the dynamical age-size relation determined in the LMC by Ambrocio-Cruz et al. (2016). 
Doing that we make the assumption that the dynamical evolution of the extraplanar H{\,\sc{ii}} regions of NGC 4254
is similar to that of typical disc H{\,\sc{ii}} regions despite the conditions of the surrounding medium (gas density and temperature) are probably 
significantly different than those generally encountered in the interstellar medium of late-type galaxies (Tonnesen \& Bryan 2012). 
For a fair comparison we need to measure the size of the H{\,\sc{ii}} regions as consistently as possible with Ambrocio-Cruz et al. (2016).

\begin{figure}
   \centering
   \includegraphics[width=9cm]{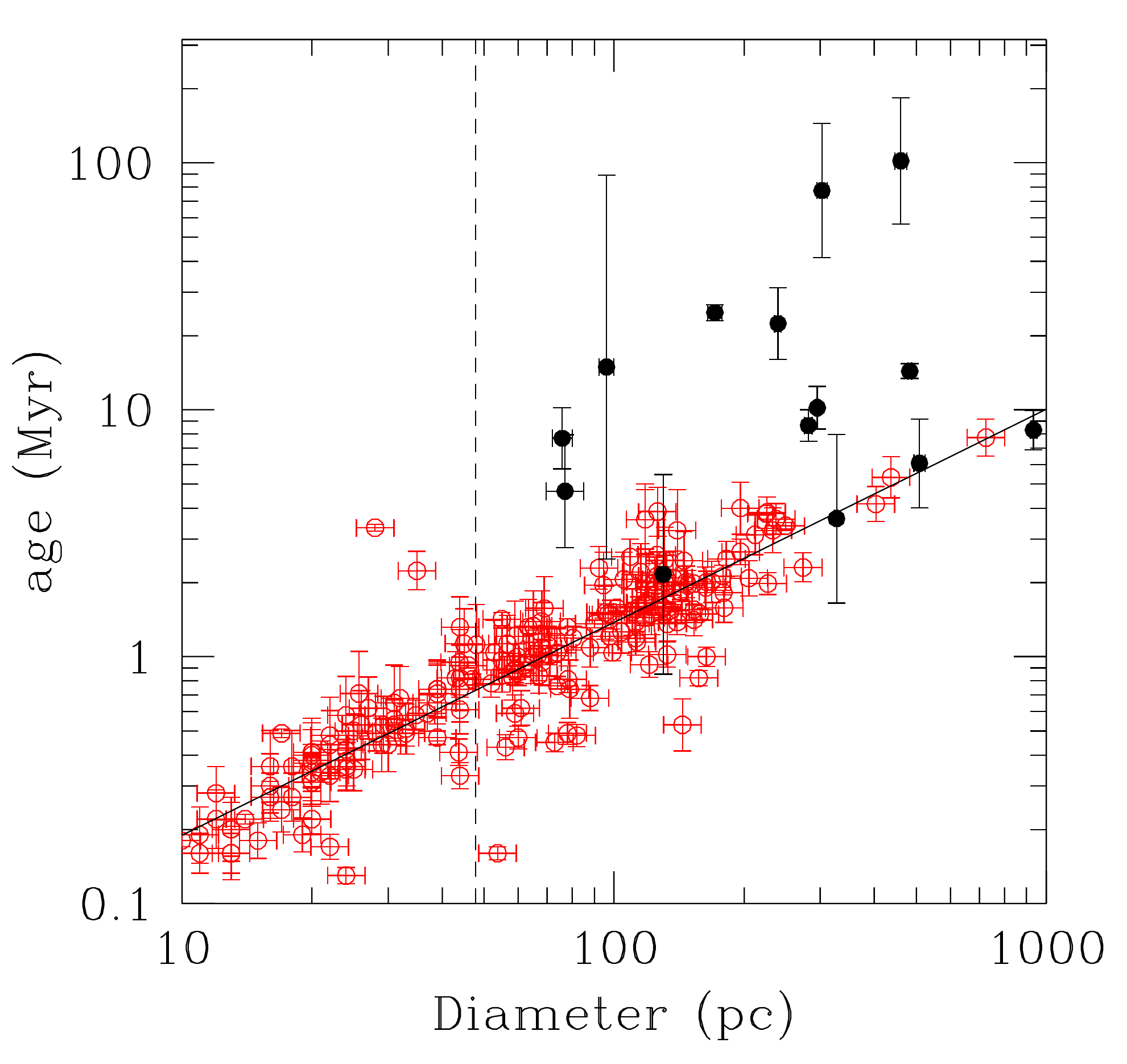}
   \caption{The dynamical age versus size relation derived for H{\,\sc{ii}} regions in the LMC (red symbols, Ambrocio-Cruz et al. 2016) is compared to the age versus size relation
   derived from SED fitting for the extraplanar H{\,\sc{ii}} regions in NGC 4254 (black filled dots). The vertical dashed line indicates the limiting seeing of the $r$- and 
   narrow-band images. The black solid line shows the best fit for H{\,\sc{ii}} regions in the LMC: log(age) = 0.86($\pm$0.03) $\times$ log(Diameter) - 1.58($\pm$0.06), $R^2$ = 0.88}
   \label{agesize}%
   \end{figure}

We measure the size of the few pointlike, symmetric H{\,\sc{ii}} regions of the sample (14/60 objects) 
with a single counterpart within the selected UV region. This last condition is required to avoid H{\,\sc{ii}} regions formed by distinct associations of newly formed stars,
for which a measure of a diameter might be meaningless.
This measurement is performed using the GALFIT code (Peng et al. 2002, 2010) run on the narrow-band H$\alpha$ image 
by assuming a Gaussian profile for the H{\,\sc{ii}} region and using an empirical PSF image derived from 138 non-saturated stars in the narrow-band image of NGC 4254.
We adopt as diameter of the H{\,\sc{ii}} region the diameter including 80\%~ of the total flux. When $\sigma$ is the width parameter of the GALFIT Gaussian model,
the diameter including 80\%~ of the total flux is $Diameter$ = 1.3 $\times$ $\sigma$.
The diameters of these H{\,\sc{ii}} regions are given in Table \ref{cigale_out}.
We then compare the age-size relation of the extraplanar H{\,\sc{ii}} of NGC 4524 (where here the
age is the one derived from SED fitting) to the dynamical age versus size relation of H{\,\sc{ii}} regions in the LMC in Fig. \ref{agesize}\footnote{The diameters of the H{\,\sc{ii}}
regions in the LMC are isophotal. Given the small size of the H{\,\sc{ii}} region on the image, and the unknown surface brightness limit used in Ambrocio-Cruz et al. (2016), we 
decided to use the $Diameter$ definition given above as a representative measure of the size of these extraplanar H{\,\sc{ii}} regions. We caution that part of the
systematic misalignment observed in Fig. \ref{agesize} can be due to the different definition of the size used for the two sets of data.}. We also notice that none of these regions
is an outlier in Fig. \ref{FUVNUVHa}.

Despite a possible large uncertainty on the measure of the diameter of the H{\,\sc{ii}} regions due to the size close to the seeing limit and to a possible
blending of multiple star clusters, Figure \ref{agesize} shows that the age derived from the SED fitting analysis is larger than the dynamical age 
expected for H{\,\sc{ii}} regions of similar size. This suggests that the ages derived from CIGALE are probably upper limits, as indeed 
expected given the limits of the stellar population synthesis codes to derive ages younger than a few million years.

\section{Discussion}

\subsection{Detection of the stripped material}

The first result of the present analysis is a further confirmation that the detection of the material stripped from a perturbed galaxy in rich
environments requires multifrequency observations to trace the different gas, dust and stellar components. There are, indeed, several
examples where the stripped material is mainly cold HI gas not associated with any process of star formation (Chung et al. 2007; 
Boissier et al. 2012; Scott et al. 2012). This gas can also be in the cold molecular phase whenever giant molecular clouds associated with star forming regions are present 
(Verdugo et al. 2015; Jachym et al. 2017), although this gas phase is not always detected (Jachym et al. 2013), or in warm H$_2$ (Sivanandam et al. 2014). 
Some of the tails are also detected as diffuse structures in H$\alpha$
(ionised gas; Gavazzi et al. 2001; Yoshida et al. 2002; Cortese et al. 2006; Yagi et al. 2007, 2010, 2017; Sun et al. 2007; Kenney et al. 2008; Fumagalli et al. 2014;
Fossati et al. 2012, 2016; Zhang et al. 2013; Boselli \& Gavazzi 2014), in X-rays (Sun et al. 2006, 2007, 2010), or
in radio continuum (Gavazzi \& Jaffe 1985, 1987; Gavazzi et al. 1995).
Others are detected mainly in UV and optical bands because of the presence of compact stellar regions formed after the interaction, 
as observed in the wake of IC 3418 in Virgo (Hester et al. 2010; Fumagalli et al. 2011a; Kenney et al. 2014) 
or the jellyfish galaxies of Poggianti et al. (2017). 

In the case of NGC 4254, the stripped material has been detected in HI (Phookun et al. 1993;
Haynes et al. 2007), while the compact H{\,\sc{ii}} regions are clearly visible only in the GALEX UV bands, or in the deep H$\alpha$ image, but only within $\lesssim$ 20 
kpc from the edge of the galaxy disc, while they are totally absent in the outer HI tail. Because of their very young age, they 
are almost undetected in the deep NGVS optical bands. NGC 4254 would thus be absent in a selection of perturbed galaxies such as the one done in the optical $B$-band 
by Poggianti et al. (2017) (GASP survey). 

\subsection{Impact of a stochastic IMF on the dating of the H{\,\sc{ii}} regions}

The analysis indicates that the observed regions have H$\alpha$ luminosities of 10$^{37}$ $\lesssim$ $L(H\alpha)$ $\lesssim$ 10$^{39}$ 
erg s$^{-1}$, which is typical of giant H{\,\sc{ii}} regions in the Milky Way or in nearby galaxies, and ages $\lesssim$ 100 Myr. 
Given their luminosity and stellar mass ($M_{star}$ $\gtrsim$ 10$^4$ M$_{\odot}$), these H{\,\sc{ii}} regions straddle the boundary of mass 
at which stochastic sampling of the IMF starts to play a role (Fumagalli et al. 2011b; da Silva et al. 2012; Cervino et al. 2013). 
At this luminosity, the H{\,\sc{ii}} regions should include $\sim$ 1000 stars (Lee et al. 2011) and quite a few ionising objects 
(Elmegreen 2000; Koda et al. 2012). Nevertheless, $M_{star}$ $\simeq$ 10$^4$ M$_{\odot}$ is close to the limit where the undersampling 
of the IMF starts to be important for the determination of the ionising stellar component
using population synthesis models (Cervino et al. 2003; Cervino \& Luridiana 2004; da Silva et al. 2012; Cervino et al. 2013).
If the IMF is undersampled, however, the observed H$\alpha$ luminosity at constant age is on average less than what predicted 
by our fully-sampled population synthesis models. Thus, the ages of the H{\,\sc{ii}} regions may be on average overestimated 
for the smallest H{\,\sc{ii}} regions. This effect would not alter our main conclusion that the extraplanar H{\,\sc{ii}} regions of 
NGC 4254 are young ($\lesssim$ 100 Myr).

\subsection{Star formation process within extraplanar H{\,\sc{ii}} regions}

IFU observations of the stripped gas of cluster galaxies indicate that star formation occurs only whenever the velocity dispersion 
in the tail is sufficiently low ($\sigma$ $\sim$ 25-50 km s$^{-1}$; Fumagalli et al. 2014; Fossati et al. 2016; Consolandi et al. 2017) 
to allow the formation of giant molecular clouds. These velocities dispersions are comparable to those observed 
over the disc of the galaxy by Wilson et al. (2011). ALMA observations are necessary to detect the molecular gas in these compact regions
and derive its typical column density and velocity dispersion. Indeed, none of the few H{\,\sc{ii}} regions mapped with HERA 
are detected in the $^{12}$CO(2-1) line (Leroy et al. 2009). This is not surprising if the extraplanar star forming regions 
of NGC 4254 have been formed within giant molecular clouds (GMC) having properties similar to those observed within the Milky Way or in nearby 
spiral galaxies, e.g. have masses in the range 5 $\times$ 10$^4$ $\lesssim$ $M_{GMC}$ $\lesssim$  5 $\times$ 10$^6$ M$_{\odot}$,
sizes 5 $\lesssim$ $R_{GMC}$ $\lesssim$ 30 pc, and thus gas column densities $\Sigma (H_2)$ $\simeq$ 170 M$_{\odot}$ pc$^{-2}$
and volume densities 50 $\lesssim$ $n(H_2)$ $\lesssim$ 500 cm$^{-3}$ (Solomon et al. 1987; Engargiola et al. 2003).
At a distance of 16.5 Mpc the filling factor of these typical giant molecular clouds 
is $\simeq$ 10\%, thus the total CO emission is sufficiently diluted not to be detected within a 13 arcsec beam. Furthermore, the CO emission
should also be low because of the expected low metallicity of the gas.

At these column densities, the typical collapse time (free fall time) for a cloud with a spherically-symmetric distribution of mass is:

\begin{equation}
{t_{ff} = \sqrt{\frac{3\pi}{32G\rho}} \simeq \rm{5 Myr}}
\end{equation} 

\noindent 
where $\rho$ is the mean density of the gas. This timescale can be compared with the time since the first interaction of the galaxy
with a nearby companion, that dynamical simulations place between 280 Myr (Vollmer et al. 2005) and 
750 Myr ago (Duc \& Bournaud 2008), and with the typical age of the stellar populations of the H{\,\sc{ii}} regions ($\lesssim$ 100 Myr). 
The stripped gas must have reached the typical column density of GMC in $\lesssim$ 40 Myr (see next section) to produce the observed extraplanar 
H{\,\sc{ii}} regions. The densest regions detected at the VLA in the HI 
tail have column densities of $\Sigma(HI)$ $\simeq$ 10$^{19}$ cm$^{-2}$ (Fig. \ref{HI}; Chung et al. 2009) and are 
unresolved at the $\sim$ 30 arcsec beam resolution. Assuming a filling factor of $\sim$ 10\%, and a clumpy distribution, it is possible
that the HI gas reaches column densities similar to those encountered in the outer discs of late-type galaxies producing molecular clouds 
where star formation takes place. Indeed, star formation seems to follow the typical Schmidt-Kennicutt relation in tidal debries (Boquien et al. 2011).
However, the physical conditions of the stripped gas within a hot intergalactic medium (gravitational potential, pressure, temperature, turbulence) 
change significantly with respect to the typical ISM in galactic discs, and it is still unclear why in these outer regions the gas collapses to form giant H{\,\sc{ii}}
regions while at further distances from the galaxy along the tail or in other similar objects within the same cluster it does not (e.g. NGC 4569, Boselli et al. 2016a).
In particular, in these extraplanar H{\,\sc{ii}} regions the lack of a dominant galactic potential well should reduce the hydrostatic pressure of the gas, 
and thus prevent the formation of molecular gas (Blitz \& Rosolowsky 2006). However, turbulence and shock fronts produced by the gravitational perturbation or
by the ram pressure that the galaxy is suffering while entering into the cluster can compress
the gas and lead to H$_2$ formation (Tonnesen \& Bryan 2010, 2012). In NGC 4254 the stripping of the gas seems to occur mostly edge-on.
Simulations suggest that in this geometrical configuration the gas, located in a thin disc before the interaction, gets less dispersed than in a face-on
stripping process, thus favouring the formation of molecular clouds (Roediger \& Br\"uggen 2006).

\begin{figure}
   \centering
   \includegraphics[width=9cm]{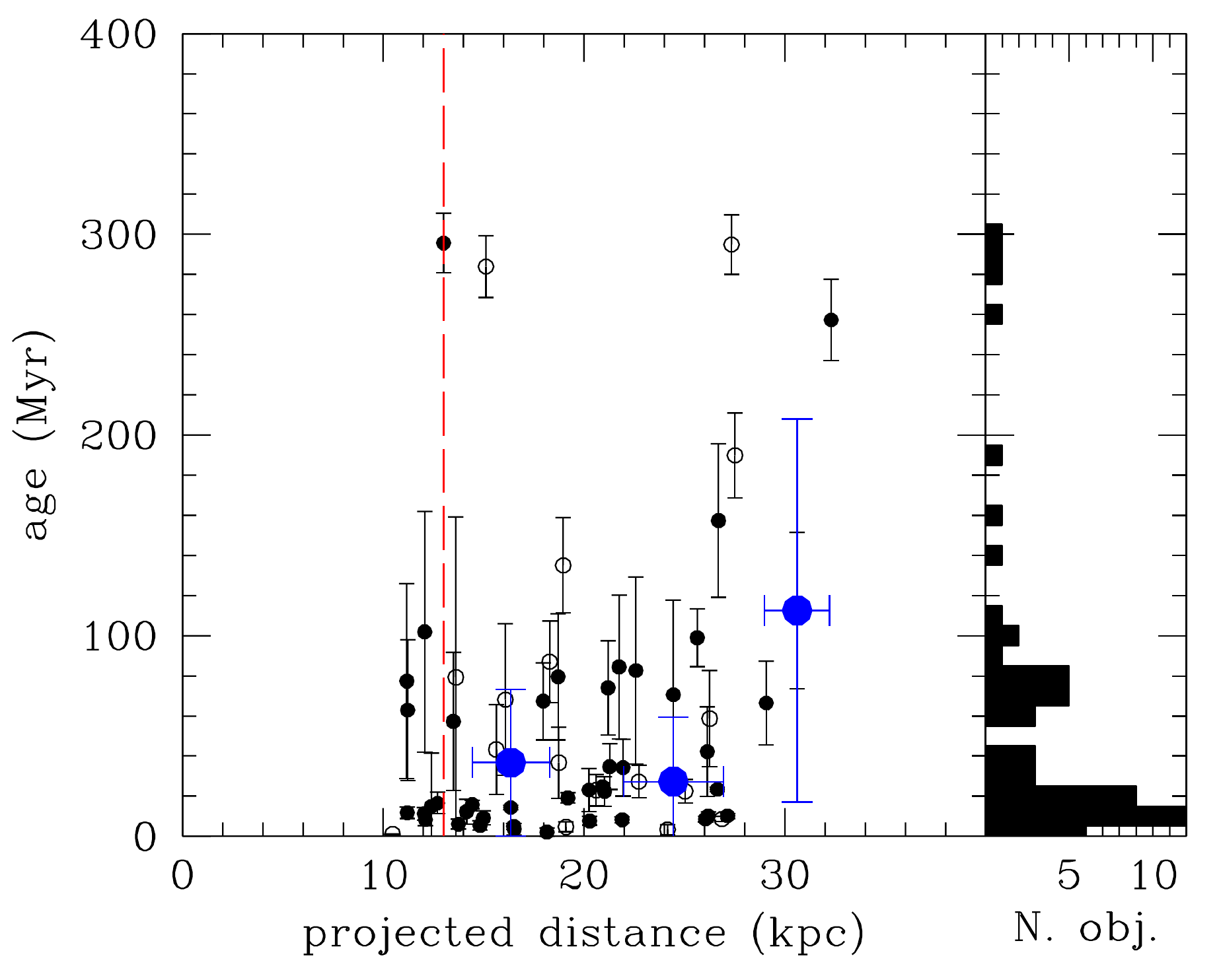}
   \caption{Left panel: relationship between the age of the different H{\,\sc{ii}} regions as derived from CIGALE (in Myr) 
   and their projected distance from the nucleus of NGC 4254 (in kpc). Filled dots indicates the H{\,\sc{ii}} regions where the SED fitting gives reduced $\chi^2_r$ $<$ 6,
   empty dots $\chi^2_r$ $\geq$ 6. Blue big filled dots indicate the median values within different distance bins.
   The red long-dashed vertical line indicates the 23.5 mag arcsec$^{-2}$ $i$-band isophotal radius of the galaxy given in Cortese et al. (2012b). Right panel:
   age distribution of all the H{\,\sc{ii}} regions.
 }
   \label{distance}%
   \end{figure}

Figure \ref{distance} shows the relation between the age of the stellar population of the different H{\,\sc{ii}} regions and their projected distance from the
nucleus of the galaxy. There is no obvious trend between the two quantities, implying that there is no strong differential evolution of these
regions after the perturbation. We recall, however, that all these regions are located within $\simeq$ 35 kpc in projected distance from the galaxy nucleus,
corresponding to $\simeq$ 2 optical radii of the galaxy. This small distance range probably corresponds to a small dynamic range 
in stellar age which is difficult to resolve with a simple SED fitting analysis, particularly with a sample characterised by a low detection rate 
at several wavelengths. 

\subsection{Fate of the stripped material}

Considering the line-of-sight motion of NGC 4254 within the cluster we can estimate the time necessary to travel $\sim$ 30 kpc, the
mean projected distance of the H{\,\sc{ii}} regions, from the galaxy. 
NGC 4254 is a member of the northern cloud of the Virgo cluster (Gavazzi et al. 1999) and has
a relative velocity of $\sim$ 1400 km s$^{-1}$ with respect to the cluster centre. Assuming that the galaxy is moving on the plane of the sky
with the typical velocity dispersion of the cluster ($\simeq$ 800 km s$^{-1}$; Boselli et al. 2014b), it would have traveled $\sim$ 30 kpc in $\sim$ 40 Myr. In the same
time the galaxy would have crossed $\sim$ 55 kpc along the line of sight. This rough estimate suggests that the furthest H{\,\sc{ii}} regions are at $R_{HII}$ $\sim$ 60 kpc
from the nucleus of the galaxy.
At these distance these H{\,\sc{ii}} regions are still gravitationally bound to NGC 4254. Indeed, the truncation radius $R_T$ of the galaxy, which lies
at a projected distance $R_{M87}$ $\sim$ 1 Mpc from the core of the cluster (M87), is (Binney \& Tremaine 2008):

\begin{equation}
{R_T = R_{M87} \times \bigg(\frac{M_{NGC4254,DM}}{3 \times M_{Virgo,DM}}\bigg)^{1/3}}
\end{equation} 

\noindent
where $M_{NGC 4254,DM}$ is the total (dark matter) mass of NGC 4254 and $M_{Virgo,DM}$ that of the Virgo cluster. 
Assuming $M_{Virgo,DM}$ = 4 $\times$ 10$^{14}$ M$_{\odot}$
(McLaughlin 1999) and $M_{NGC 4254,DM}$ = 10$^{12}$ M$_{\odot}$ as derived from Behroozi et al. (2010) for a stellar mass of NGC 4254 of $M_{NGC 4254,star}$ = 2.4 $\times$ 10$^{10}$ M$_{\odot}$,
the truncation radius of NGC 4254 is $R_T$ $\simeq$ 95 kpc. These regions are within the truncation radius of the galaxy and might thus fall back on the disc.
However, the HI observations of Phookun et al. (1993) suggest that the HI clouds associated with these regions have a $\sim$ 200 km s$^{-1}$
lower recessional velocity along the line of sight than that of the galaxy (see Fig. \ref{HI}), corresponding to $\sim$ 340 km s$^{-1}$ in a
3-D space. This value is close to the escape velocity from a spherically symmetric solid body given by the relation:

\begin{equation}
{v_e = \sqrt{\frac{2GM_{NGC4254,DM}}{R_{HII}}}}
\end{equation}
 
\noindent
that for the furthest H{\,\sc{ii}} regions ($R_{HII}$ $\sim$ 60 kpc) is $\sim$ 390 km s$^{-1}$. Furthermore, the galaxy moving towards the cluster centre is 
undoubtly undergoing a strong ram pressure event as clearly seen in the asymmetric distribution of the radio continuum emission (Kantharia et al. 2008) and of the
HI gas over its disc, with an extended tail in the 
northern direction (Phookun et al. 1993; Kantharia et al. 2008). This external pressure might have favoured the formation of giant molecular clouds (Blitz \& Rosolowsky 2006), 
although some clouds might have been stripped from the wester spiral arm during the interaction. Here, indeed, ram pressure is made more efficient than 
elsewere in the disc because of the combined effect of the edge-on motion of the galaxy within the cluster and its clockwise rotation.
The HI gas blobs associated to the extraplanar H{\,\sc{ii}} regions are suffering the same ram pressure event affecting the galaxy, as indeed suggested by their lower recessional velocity
with respect to NGC 4254 ($\simeq$ 200 km s$^{-1}$, see Fig. \ref{HI}), preventing them to fall back onto the stellar disc. 
Given these considerations, while keeping in mind the large uncertainties on all the rough calculations made above, 
it is possible that some of the extraplanar H{\,\sc{ii}} regions will become free-floating objects within the cluster, as indeed indicated by the dynamical simulations 
of Duc \& Bournaud (2008). In that case, these compact stellar objects can contribute to the diffuse intracluster light of Virgo (Mihos et al. 2005, 2009, 2017).
Because of their compact nature and their single-age stellar populations, these regions can be the progenitors of globular clusters, diffuse star clusters (DSC), and ultra compact dwarf
galaxies (UCD), mainly observed around the dominant massive elliptical galaxies in nearby clusters (Peng et al. 2006; Liu et al. 2015, 2016). Their stellar masses 
(see Table \ref{cigale_out} and Fig. \ref{Hadist}) are less than those of typical UCDs and globular clusters, but these last have been formed several Gyr ago, when
the typical gas content of galaxies was significantly larger than in similar objects in the local universe. It is thus plausible that 
high-speed fly-by encounters of gas-rich galaxies in forming clusters at high redshift might have produced more massive compact objects than those 
observed around NGC 4254, giving birth to globular clusters, DSCs, and UCDs such as those observed in the local universe.
The properties of the extraplanar H{\,\sc{ii}} regions of NGC 4254 in terms of stellar mass are comparable to those observed in other Virgo cluster compact sources
formed after a gravitational perturbation between galaxies (M 49 - VCC 1249; Arrigoni-Battaia et al. 2012) or by ram pressure stripping (IC 3418 - Hester et al. 2010;
Fumagalli et al. 2011a; Kenney et al. 2014). They are remarkably similars to those of the ultra-compact high velocity cloud AGC 226067, a possible stripping 
remnant in the Virgo cluster, characterised by a single stellar population with an age of $\simeq$ 7-50 Myr (Sand et al. 2017).

\section{Conclusion}

Very deep H$\alpha$ images obtained during the VESTIGE survey of the Virgo cluster ($f(H\alpha)$ $\sim$ 4$\times$10$^{-17}$ erg s$^{-1}$ cm$^{-2}$ for point souces at 5$\sigma$ and 
$\Sigma(H\alpha)$ $\sim$ 2$\times$10$^{-18}$ erg s$^{-1}$ cm$^{-2}$ arcsec$^{-2}$ for extended sources at 1$\sigma$), combined with UV GALEX and optical NGVS
data, revealed the presence of 60 compact (70-500 pc) star forming regions outside the optical disc of the massive spiral galaxy NGC 4254 (M99). 
These regions are located along the tail of HI gas harassed from the galaxy after a gravitational interaction with another Virgo cluster member
that simulations indicate occurred between 280 and 750 Myr ago (Vollmer et al. 2005; Duc \& Bournaud 2008).
The analysis of their stellar populations, as well as their physical sizes, consistently indicate that these regions are young 
($\lesssim$ 100 Myr). These observations offer new observational constraints for models and simulations of star formation in the 
stripped gas of cluster galaxies. Consistent with other observations of perturbed galaxies, these data indicate that the process of star formation
outside the disc of galaxies is very episodic, lasts for a few Myr, and occurs only in compact regions. The velocity dispersion of the gas in these compact regions
is expected to be sufficiently low to allow matter to collapse and form 
giant molecular clouds able to shield the gas from the external heating due to the hot intracluster medium. 
A complete multiphase modelling of the star formation process within these extreme regions requires more observational constraints on the different gas phases (HI,
H$_2$, X-rays), as well as high-resolution spectroscopy for the determination of their kinematic and physical properties.

\begin{acknowledgements}

We thank the anonymous referee for useful comments on the manuscript.
We are grateful to M. Haynes for providing us with the HI data of NGC 4254, to C. Spengler for the use of GALFIT, and to 
the whole CFHT team who assisted us in the preparation and in the execution of the observations and in the calibration and data reduction: 
Todd Burdullis, Daniel Devost, Bill Mahoney, Nadine Manset, Andreea Petric, Simon Prunet, Kanoa Withington.
We acknowledge financial support from "Programme National de Cosmologie and Galaxies" (PNCG) funded by CNRS/INSU-IN2P3-INP, CEA and CNES, France, and from
"Projet International de Coop\'eration Scientifique" (PICS) with Canada funded by the CNRS, France.
This research has made use of the NASA/IPAC Extragalactic Database (NED) 
which is operated by the Jet Propulsion Laboratory, California Institute of 
Technology, under contract with the National Aeronautics and Space Administration
and of the GOLDMine database (http://goldmine.mib.infn.it/) (Gavazzi et al. 2003).
MB was supported by MINEDUC-UA projects, code ANT 1655 and ANT 1656.
MS acknowledges support from the NSF grant 1714764 and the Chandra Award GO6-17111X.
MF acknowledges support by the science and technology facilities council [grant number ST/P000541/1].

\end{acknowledgements}

\begin{landscape}
\begin{table}
\caption{Flux densities of the H{\,\sc{ii}} regions. }
\resizebox{22cm}{!}{
\begin{tabular}{cccccccccccccccc}
\hline
\hline
Region  & R.A.(2000) 	& Dec		& $r_{HII}$	& $LyC^a$	& $FUV$	& $NUV$	& $u$	& $g$	&  $i$	& $z$	& 3.6$\mu$m	& 4.5$\mu$m	& 5.8$\mu$m	& 8.0$\mu$m	& 24$\mu$m \\
	& h m s		& $^o$ $'$ $"$	& $"$		& $\mu$Jy	& $\mu$Jy	& $\mu$Jy	& $\mu$Jy	& $\mu$Jy	& $\mu$Jy	& $\mu$Jy	& $\mu$Jy		& $\mu$Jy		& $\mu$Jy		& $\mu$Jy		& $\mu$Jy\\
\hline
           1 &12:18:43.617&+14:23:24.13&6.52&	       196.3   $\pm$	   1.8    &    2.1$\pm$0.2    &     2.4$\pm$0.3    &		      7.1$\pm$   2.3  &        $<$	   5.4  	   &	   	 $<$	     9.0	      &       $<$	 11.0		   &	   $<$         7.0		&		   16.0$\pm$   5.0   &  		56.0$\pm$  15.0   &		    171.0$\pm$  43.0   &       $<$	  74.0  	    \\ 
           2 &12:18:42.478&+14:23:12.26&5.76&	        26.7   $\pm$	   1.6    &    4.5$\pm$0.6    &     5.2$\pm$0.9    &	  $<$	      7.5	      &        $<$	  17.2  	   &	   	 $<$	    31.7	      &       $<$	 42.1		   &		     128.0$\pm$  25.0	&		  111.0$\pm$  18.0   &       $<$	45.0		  &	  $<$	    169.0	       &       $<$	 121.0  	    \\ 
           3 &12:18:47.554&+14:21:37.30&4.36&	        33.4   $\pm$	   1.2    &    3.4$\pm$0.3    &     3.7$\pm$0.3    &		      6.7$\pm$   0.8  & 		  11.0$\pm$   1.4  &	   		    11.1$\pm$	3.0   & 		 12.9$\pm$   3.8   &	   $<$         6.0		&		   17.0$\pm$   5.0   &       $<$	12.0		  &		     83.0$\pm$  25.0   &       $<$	  34.0  	    \\ 
           4 &12:18:42.664&+14:22:42.60&6.38&	        22.1   $\pm$	   1.3    &    4.9$\pm$0.6    &     5.5$\pm$0.8    &		      7.9$\pm$   2.4  &        $<$	   5.4  	   &	   	 $<$	     9.0	      &       $<$	 10.7		   &	   $<$        17.0		&	$<$	   13.0 	     &       $<$	36.0		  &	  $<$	    146.0	       &       $<$	  72.0  	    \\ 
           5 &12:18:39.060&+14:22:46.56&5.46&	        35.6   $\pm$	   0.8    &    4.5$\pm$0.4    &     4.4$\pm$0.4    &		      6.5$\pm$   1.7  &        $<$	   3.5  	   &	   	 $<$	     5.4	      &       $<$	  7.0		   &	   $<$         5.0		&		   21.0$\pm$   5.0   &       $<$	24.0		  &	  $<$	     45.0	       &       $<$	  54.0  	    \\ 
           6 &12:18:38.200&+14:22:30.09&6.37&	        15.9   $\pm$	   1.0    &    2.0$\pm$0.3    &     1.7$\pm$0.4    &	  $<$	      1.7	      &        $<$	   3.7  	   &	   	 $<$	     6.9	      &       $<$	  6.0		   &	   $<$         9.0		&	$<$	    8.0 	     &       $<$	32.0		  &	  $<$	     55.0	       &       $<$	  38.0  	    \\ 
           7 &12:18:32.956&+14:23:41.26&6.94&	        10.2   $\pm$	   1.9    &    8.7$\pm$0.6    &    10.2$\pm$0.6    &		     11.9$\pm$   1.2  & 		  17.8$\pm$   2.4  &	   		    24.9$\pm$	4.3   & 		 32.2$\pm$   6.5   &		      89.0$\pm$  11.0	&		   67.0$\pm$   8.0   &       $<$	22.0		  &	  $<$	     49.0	       &       $<$	  43.0  	    \\ 
           8 &12:18:31.743&+14:24:02.94&6.14&	        17.0   $\pm$	   2.5    &    5.4$\pm$0.4    &     7.0$\pm$0.4    &		      5.1$\pm$   0.8  &        $<$	   1.6  	   &	   	 $<$	     7.1	      &       $<$	  5.3		   &	   $<$        16.0		&	$<$	   13.0 	     &       $<$	27.0		  &	  $<$	     29.0	       &       $<$	  33.0  	    \\ 
           9 &12:18:36.965&+14:22:10.97&5.16&	        19.3   $\pm$	   1.2    &    2.9$\pm$0.3    &     3.6$\pm$0.3    &		      3.2$\pm$   0.6  & 		   3.3$\pm$   1.0  &	   	 $<$	     3.0	      &       $<$	  2.9		   &	   $<$         6.0		&	$<$	    7.0 	     &       $<$	15.0		  &	  $<$	     23.0	       &       $<$	  24.0  	    \\ 
          10 &12:18:38.714&+14:21:46.03&6.86&	         9.3   $\pm$	   1.7    &    8.5$\pm$0.6    &    10.6$\pm$0.7    &		     14.0$\pm$   1.5  & 		  15.6$\pm$   2.2  &	   	 $<$	     5.6	      &       $<$	  8.4		   &	   $<$        17.0		&	$<$	   16.0 	     &       $<$	24.0		  &	  $<$	    132.0	       &       $<$	  50.0  	    \\ 
          11 &12:18:37.016&+14:21:56.62&8.67&	        14.2   $\pm$	   3.3    &   18.8$\pm$1.0    &    19.5$\pm$1.1    &		     24.7$\pm$   2.0  & 		  25.4$\pm$   2.7  &	   	 $<$	     9.7	      &       $<$	  9.3		   &	   $<$        21.0		&	$<$	   18.0 	     &       $<$	29.0		  &	  $<$	     57.0	       &       $<$	  59.0  	    \\ 
          12 &12:18:34.395&+14:21:04.64&7.03&	        42.3   $\pm$	   1.2    &   11.0$\pm$0.7    &    11.0$\pm$0.8    &		     10.9$\pm$   1.6  & 		  11.3$\pm$   1.8  &	   		    10.3$\pm$	2.9   &       $<$	  5.7		   &	   $<$        12.0		&	$<$	    7.0 	     &       $<$	25.0		  &	  $<$	     60.0	       &       $<$	  31.0  	    \\ 
          13 &12:18:33.383&+14:21:02.78&6.75&	        14.4   $\pm$	   1.0    &    6.1$\pm$0.6    &     6.0$\pm$0.7    &		      5.6$\pm$   1.1  &        $<$	   2.1  	   &	   		     6.1$\pm$	1.6   &       $<$	  6.6		   &	   $<$         5.0		&	$<$	    9.0 	     &       $<$	21.0		  &	  $<$	     35.0	       &       $<$	  34.0  	    \\ 
          14 &12:18:38.171&+14:20:50.76&4.49&	         4.7   $\pm$	   0.8    &    0.7$\pm$0.2    &     0.9$\pm$0.2    &	  $<$	      0.5	      &        $<$	   0.8  	   &	   	 $<$	     2.3	      &       $<$	  2.4		   &	   $<$         4.0		&	$<$	    4.0 	     &       $<$	10.0		  &	  $<$	     30.0	       &       $<$	  18.0  	    \\ 
          15 &12:18:30.260&+14:22:31.69&4.45&	         9.7   $\pm$	   0.8    &    2.7$\pm$0.2    &     3.3$\pm$0.2    &		      2.5$\pm$   0.3  & 		   2.7$\pm$   0.4  &	   	 $<$	     0.9	      &       $<$	  1.8		   &	   $<$         3.0		&	$<$	    4.0 	     &       $<$	13.0		  &	  $<$	     21.0	       &       $<$	  22.0  	    \\ 
          16 &12:18:35.531&+14:21:18.33&11.1&	    $<$ 15.7        	          &   10.1$\pm$0.8    &    10.1$\pm$1.9    &	  $<$	     33.0	      &        $<$	  68.7  	   &	   	 $<$	   110.0	      &       $<$	 88.6		   &	   $<$        27.0		&	$<$	   23.0 	     &       $<$	20.0		  &	  $<$	    200.0	       &       $<$	  36.0  	    \\ 
          17 &12:18:32.664&+14:26:42.99&13.0&	    $<$  6.7        	          &   15.5$\pm$1.0    &    16.9$\pm$1.3    &	  $<$	     10.1	      &        $<$	  15.0  	   &	   	 $<$	    25.5	      &       $<$	 37.8		   &	   $<$        62.0		&	$<$	   47.0 	     &       $<$	94.0		  &	  $<$	     64.0	       &       $<$	 133.0  	    \\ 
          18 &12:18:34.726&+14:23:31.23&5.46&	         6.7   $\pm$	   1.3    &    4.7$\pm$0.4    &     5.8$\pm$0.4    &		      6.6$\pm$   1.2  & 		   9.0$\pm$   2.6  &	   	 $<$	     4.3	      &       $<$	  6.4		   &		      26.0$\pm$   8.0	&		   22.0$\pm$   5.0   &       $<$	13.0		  &	  $<$	     22.0	       &       $<$	  82.0  	    \\ 
          19 &12:18:37.598&+14:24:15.38&8.31&	       151.0   $\pm$	   3.9    &    8.1$\pm$1.2    &     9.5$\pm$2.1    &	  $<$	     15.2	      &        $<$	  40.4  	   &	   	 $<$	    65.6	      &       $<$	 91.6		   &	   $<$        66.0		&	$<$	   45.0 	     &       $<$       120.0		  &		   1110.0$\pm$ 168.0   &       $<$	 215.0  	    \\ 
          20 &12:18:27.741&+14:23:57.16&7.18&	    $<$  1.9        	          &    9.5$\pm$0.5    &    12.2$\pm$0.6    &		     19.9$\pm$   1.7  & 		  39.3$\pm$   2.6  &	   		    59.8$\pm$	5.1   & 		 58.4$\pm$   5.9   &	   $<$        18.0		&	$<$	   16.0 	     &       $<$	25.0		  &	  $<$	     38.0	       &       $<$	  60.0  	    \\ 
          21 &12:18:32.182&+14:21:32.91&9.94&	        29.2   $\pm$	   3.3    &   14.9$\pm$0.8    &    17.0$\pm$1.0    &		     20.6$\pm$   1.8  & 		  31.4$\pm$   2.2  &	   		    62.9$\pm$	6.9   & 		 70.0$\pm$  15.2   &		     121.0$\pm$  17.0	&		  111.0$\pm$  16.0   &  	       128.0$\pm$  42.0   &		    199.0$\pm$  54.0   &       $<$	 120.0  	    \\ 
          22 &12:18:45.170&+14:22:56.25&7.13&	        12.5   $\pm$	   2.6    &    6.9$\pm$1.4    &     7.8$\pm$2.0    &	  $<$	      8.5	      &        $<$	  18.2  	   &	   	 $<$	    31.2	      &       $<$	 46.6		   &	   $<$        33.0		&	$<$	   24.0 	     &       $<$	80.0		  &	  $<$	    150.0	       &       $<$	 198.0  	    \\ 
          23 &12:18:47.413&+14:22:02.02&7.20&	        30.5   $\pm$	   3.0    &    9.7$\pm$0.7    &    11.4$\pm$0.8    &		     14.5$\pm$   3.3  &        $<$	   7.3  	   &	   		    57.2$\pm$  16.5   &       $<$	 20.8		   &		      61.0$\pm$  18.0	&		   52.0$\pm$  13.0   &       $<$	30.0		  &	  $<$	    221.0	       &       $<$	 123.0  	    \\ 
          24 &12:18:51.095&+14:22:30.27&7.15&	    $<$  2.2        	          &    5.2$\pm$0.9    &     7.4$\pm$1.7    &	  $<$	      9.8	      &        $<$	  17.8  	   &	   	 $<$	    25.4	      &       $<$	 36.0		   &	   $<$        16.0		&		   37.0$\pm$  11.0   &       $<$	51.0		  &	  $<$	    112.0	       &       $<$	 121.0  	    \\ 
          25 &12:18:48.241&+14:22:41.21&5.90&	    $<$  3.5        	          &    4.4$\pm$0.9    &     5.9$\pm$1.1    &	  $<$	      7.0	      &        $<$	  15.1  	   &	   	 $<$	    23.1	      &       $<$	 32.5		   &	   $<$        26.0		&	$<$	   17.0 	     &       $<$	63.0		  &	  $<$	    183.0	       &       $<$	 140.0  	    \\ 
          26 &12:18:46.564&+14:22:46.86&5.15&	    $<$  1.7        	          &    5.4$\pm$0.8    &     6.5$\pm$1.2    &	  $<$	      6.3	      &        $<$	  13.1  	   &	   	 $<$	    19.8	      &       $<$	 30.0		   &	   $<$        19.0		&	$<$	   14.0 	     &       $<$	60.0		  &	  $<$	    110.0	       &       $<$	 112.0  	    \\ 
          27 &12:18:40.739&+14:23:02.72&6.74&	         9.0   $\pm$	   2.0    &    3.9$\pm$0.6    &     5.7$\pm$0.8    &	  $<$	      7.3	      &        $<$	  20.1  	   &	   	 $<$	    38.6	      &       $<$	 50.0		   &	   $<$        39.0		&	$<$	   21.0 	     &       $<$	42.0		  &	  $<$	    164.0	       &       $<$	  79.0  	    \\ 
          28 &12:18:36.803&+14:23:43.90&5.44&	        31.1   $\pm$	   1.2    &    3.0$\pm$0.4    &     3.8$\pm$0.5    &	  $<$	      3.9	      &        $<$	   7.6  	   &	   	 $<$	    12.1	      &       $<$	 15.4		   &		      36.0$\pm$  11.0	&		   43.0$\pm$   8.0   &       $<$	28.0		  &		    191.0$\pm$  37.0   &       $<$	 111.0  	    \\ 
          29 &12:18:34.313&+14:23:59.06&4.94&	    $<$  1.2        	          &    2.0$\pm$0.2    &     7.1$\pm$0.4    &		      5.9$\pm$   1.4  & 		  10.3$\pm$   3.2  &	   		    30.6$\pm$	5.7   & 		 36.2$\pm$   7.7   &		     167.0$\pm$   7.0	&		  169.0$\pm$   5.0   &  	       213.0$\pm$  18.0   &		    233.0$\pm$  31.0   &       $<$	 152.0  	    \\ 
          30 &12:18:47.116&+14:21:17.90&6.81&	    $<$  2.3        	          &    3.4$\pm$0.4    &     5.5$\pm$0.5    &		      6.1$\pm$   1.0  & 		   8.3$\pm$   2.4  &	   	 $<$	     5.3	      &       $<$	  6.8		   &	   $<$        10.0		&	$<$	    7.0 	     &       $<$	17.0		  &	  $<$	     31.0	       &       $<$	  32.0  	    \\ 
          31 &12:18:42.488&+14:20:56.87&10.0&	        16.0   $\pm$	   2.5    &   18.2$\pm$1.0    &    20.7$\pm$1.2    &		     29.6$\pm$   2.7  & 		  37.9$\pm$   3.5  &	   		    48.4$\pm$	9.5   & 		 32.3$\pm$   8.4   &	   $<$        18.0		&	$<$	   18.0 	     &       $<$	29.0		  &	  $<$	    283.0	       &       $<$	 109.0  	    \\ 
          32 &12:18:37.661&+14:20:33.67&8.57&	    $<$  2.2        	          &   12.7$\pm$0.8    &    17.0$\pm$0.9    &		     17.0$\pm$   3.0  & 		  33.9$\pm$   2.2  &	   		    38.0$\pm$	7.8   & 		 49.3$\pm$   7.6   &	   $<$        19.0		&	$<$	   10.0 	     &       $<$	28.0		  &	  $<$	    146.0	       &       $<$	  59.0  	    \\ 
          33 &12:18:36.716&+14:20:14.85&7.05&	    $<$  1.6        	          &    6.3$\pm$0.4    &     9.6$\pm$0.5    &		     17.8$\pm$   3.6  & 		  31.7$\pm$   2.3  &	   		    56.7$\pm$	4.7   &       $<$	  7.4		   &	   $<$        16.0		&	$<$	    7.0 	     &       $<$	23.0		  &	  $<$	    106.0	       &       $<$	  76.0  	    \\ 
          34 &12:18:25.045&+14:22:13.88&7.25&	    $<$  2.3        	          &    7.3$\pm$0.5    &    12.8$\pm$0.7    &		     20.5$\pm$   1.6  & 		  51.0$\pm$   2.2  &	   		    79.9$\pm$	4.7   & 		 82.0$\pm$   6.3   &	   $<$        13.0		&	$<$	   20.0 	     &       $<$	30.0		  &	  $<$	     41.0	       &       $<$	  34.0  	    \\ 
          35 &12:18:25.655&+14:22:47.32&6.34&	    $<$  1.5    	          &    2.9$\pm$0.3    &     4.6$\pm$0.4    &		      5.4$\pm$   0.8  & 		   9.2$\pm$   1.2  &	   		    16.5$\pm$	3.4   & 		 20.5$\pm$   4.0   &	   $<$         6.0		&	$<$	    7.0 	     &       $<$	17.0		  &	  $<$	     24.0	       &       $<$	  36.0  	    \\ 
          36 &12:18:31.383&+14:21:57.49&7.26&	    $<$  1.4    	          &    3.5$\pm$0.3    &     5.5$\pm$0.4    &		      8.2$\pm$   0.9  & 		   7.3$\pm$   1.1  &	   	 $<$	     3.5	      &       $<$	  5.8		   &		      42.0$\pm$   7.0	&		   51.0$\pm$   8.0   &       $<$	25.0		  &	  $<$	     31.0	       &       $<$	  73.0  	    \\ 
          37 &12:18:35.001&+14:26:23.02&6.36&	    $<$  2.0        	          &    3.3$\pm$0.3    &     3.7$\pm$0.3    &	  $<$	      1.9	      &        $<$	   4.9  	   &	   	 $<$	    10.0	      &       $<$	 11.6		   &	   $<$        10.0		&	$<$	    8.0 	     &       $<$	34.0		  &	  $<$	     28.0	       &       $<$	  43.0  	    \\ 
          38 &12:18:39.944&+14:21:26.40&6.32&	    $<$  1.6        	          &    2.2$\pm$0.3    &     2.9$\pm$0.3    &	  $<$	      0.9	      &        $<$	   1.2  	   &	   	 $<$	     2.1	      &       $<$	  5.2		   &	   $<$         7.0		&	$<$	    7.0 	     &       $<$	13.0		  &	  $<$	     48.0	       &       $<$	  32.0  	    \\ 
          39 &12:18:48.075&+14:22:26.20&5.87&	         9.1   $\pm$       2.7    &    2.6$\pm$0.6    &     3.3$\pm$0.9    &	  $<$	      6.2	      &        $<$	  12.1  	   &	   	 $<$	    19.0	      &       $<$	 30.0		   &	   $<$        22.0		&	$<$	   15.0 	     &       $<$	42.0		  &	  $<$	    166.0	       &       $<$	  99.0  	    \\ 
          40 &12:18:44.091&+14:22:53.29&5.82&	         6.3   $\pm$	   1.4    &    2.9$\pm$0.6    &     3.7$\pm$0.9    &	  $<$	      3.2	      &        $<$	   7.9  	   &	   	 $<$	    14.3	      &       $<$	 23.2		   &	   $<$        17.0		&	$<$	   14.0 	     &       $<$	41.0		  &	  $<$	    107.0	       &       $<$	  93.0  	    \\ 
          41 &12:18:37.781&+14:23:58.92&5.38&	         6.9   $\pm$	   1.3    &    2.1$\pm$0.5    &     3.2$\pm$0.7    &	  $<$	      5.6	      &        $<$	  12.8  	   &	   	 $<$	    21.9	      &       $<$	 26.6		   &	   $<$        17.0		&	$<$	   13.0 	     &       $<$	42.0		  &	  $<$	     67.0	       &       $<$	  58.0  	    \\ 
          42 &12:18:38.773&+14:25:07.86&5.68&	    $<$  1.6        	          &    6.7$\pm$0.6    &    13.0$\pm$1.2    &		     59.6$\pm$   8.2  & 		 145.0$\pm$  17.6  &	   		   258.6$\pm$  32.1   & 		279.9$\pm$  39.0   &		     186.0$\pm$  24.0	&		  132.0$\pm$  17.0   &  	       212.0$\pm$  42.0   &		    720.0$\pm$  83.0   &       $<$	 181.0  	    \\ 
          43 &12:18:34.668&+14:22:07.67&5.41&	    $<$  1.1        	          &    1.4$\pm$0.2    &     2.3$\pm$0.3    &		      2.2$\pm$   0.6  & 		   4.6$\pm$   1.2  &	   		     9.3$\pm$	2.6   & 		 12.3$\pm$   3.1   &		      19.0$\pm$   6.0	&	$<$	    6.0 	     &       $<$	10.0		  &	  $<$	     17.0	       &       $<$	  63.0  	    \\ 
          44 &12:18:39.821&+14:21:57.23&7.39&	    $<$  1.1        	          &    3.1$\pm$0.4    &     6.2$\pm$0.6    &		      7.6$\pm$   2.4  &        $<$	   4.9  	   &	   	 $<$	     9.8	      &       $<$	 11.3		   &	   $<$        14.0		&	$<$	   15.0 	     &       $<$	22.0		  &	  $<$	    126.0	       &       $<$	 131.0  	    \\ 
          45 &12:18:33.428&+14:21:20.48&10.0&	    $<$  2.9        	          &    6.5$\pm$0.8    &     7.2$\pm$1.9    &	  $<$	      4.3	      &        $<$	  10.3  	   &	   	 $<$	    20.3	      &       $<$	 21.6		   &	   $<$        23.0		&	$<$	   20.0 	     &       $<$	54.0		  &	  $<$	     43.0	       &       $<$	  62.0  	    \\ 
          46 &12:18:43.437&+14:22:50.23&5.22&	         3.6   $\pm$	   1.1    &    3.3$\pm$0.5    &     3.7$\pm$0.6    &	  $<$	      2.0	      &        $<$	   4.8  	   &	   	 $<$	     8.1	      &       $<$	 11.4		   &	   $<$        13.0		&	$<$	   10.0 	     &       $<$	27.0		  &	  $<$	     92.0	       &       $<$	  53.0  	    \\ 
          47 &12:18:42.200&+14:22:11.70&5.28&	    $<$  0.7        	          &    1.6$\pm$0.3    &     2.7$\pm$0.3    &	  $<$	      1.7	      &        $<$	   3.8  	   &	   	 $<$	     7.0	      &       $<$	  8.2		   &	   $<$         5.0		&		   17.0$\pm$   5.0   &       $<$	13.0		  &	  $<$	     40.0	       &       $<$	  20.0  	    \\ 
          48 &12:18:46.838&+14:22:15.93&6.17&	    $<$  1.5        	          &    2.6$\pm$0.4    &     2.9$\pm$0.5    &	  $<$	      2.6	      &        $<$	   6.3  	   &	   	 $<$	    12.7	      &       $<$	 17.1		   &	   $<$        15.0		&	$<$	   13.0 	     &       $<$	37.0		  &	  $<$	    201.0	       &       $<$	  92.0  	    \\ 
          49 &12:18:32.262&+14:22:05.30&5.60&	    $<$  1.0    	          &    2.0$\pm$0.2    &     2.7$\pm$0.3    &	  $<$	      0.7	      &        $<$	   0.9  	   &	   	 $<$	     2.2	      &       $<$	  3.0		   &	   $<$         6.0		&	$<$	    6.0 	     &       $<$	12.0		  &	  $<$	     20.0	       &       $<$	  39.0  	    \\ 
          50 &12:18:25.457&+14:24:28.89&5.67&	    $<$  1.8        	          &    1.7$\pm$0.2    &     4.0$\pm$0.3    &		      4.7$\pm$   0.5  & 		   9.0$\pm$   0.7  &	   		    21.2$\pm$	2.0   & 		 26.9$\pm$   3.1   &	   $<$        12.0		&	$<$	   10.0 	     &       $<$	23.0		  &		    105.0$\pm$  30.0   &       $<$	  93.0  	    \\ 
          51 &12:18:38.412&+14:21:33.05&5.97&	    $<$  1.6        	          &    3.4$\pm$0.4    &     3.8$\pm$0.4    &		      2.7$\pm$   0.8  & 		   5.1$\pm$   1.3  &	   	 $<$	     3.7	      &       $<$	  5.9		   &	   $<$         9.0		&	$<$	    8.0 	     &       $<$	16.0		  &	  $<$	     91.0	       &       $<$	  31.0  	    \\ 
          52 &12:18:37.334&+14:21:22.79&5.78&	    $<$  1.4        	          &    1.9$\pm$0.3    &     2.6$\pm$0.4    &	  $<$	      0.9	      &        $<$	   1.4  	   &	   	 $<$	     3.7	      &       $<$	  4.8		   &		      26.0$\pm$   7.0	&		   31.0$\pm$   7.0   &       $<$	15.0		  &	  $<$	     91.0	       &       $<$	  89.0  	    \\ 
          53 &12:18:36.292&+14:22:48.38&4.79&	         5.4   $\pm$	   1.2    &    1.1$\pm$0.2    &     1.3$\pm$0.2    &	  $<$	      0.6	      &        $<$	   1.1  	   &	   	 $<$	     1.9	      &       $<$	  2.0		   &	   $<$         3.0		&	$<$	    3.0 	     &       $<$	10.0		  &	  $<$	     14.0	       &       $<$	  16.0  	    \\ 
          54 &12:18:31.799&+14:25:50.36&5.10&	    $<$  0.6        	          &    1.4$\pm$0.2    &     3.7$\pm$0.2    &		      4.7$\pm$   0.6  & 		   7.4$\pm$   1.1  &	   		    16.9$\pm$	2.2   & 		 18.5$\pm$   3.0   &		      23.0$\pm$   4.0	&		   25.0$\pm$   5.0   &       $<$	21.0		  &	  $<$	     14.0	       &       $<$	  20.0  	    \\ 
          55 &12:18:39.979&+14:22:47.69&6.68&	    $<$  1.4        	          &    3.0$\pm$0.4    &     3.3$\pm$0.4    &	  $<$	      1.9	      &        $<$	   5.5  	   &	   	 $<$	    11.0	      &       $<$	 15.8		   &	   $<$        10.0		&	$<$	    8.0 	     &       $<$	29.0		  &	  $<$	    121.0	       &       $<$	  37.0  	    \\ 
          56 &12:18:30.357&+14:22:07.78&5.95&	        11.2   $\pm$	   1.2    &    1.7$\pm$0.2    &     2.6$\pm$0.2    &		      1.7$\pm$   0.4  &        $<$	   0.6  	   &	   	 $<$	     2.3	      &       $<$	  3.3		   &		      34.0$\pm$   4.0	&		   47.0$\pm$   5.0   &       $<$	20.0		  &		    114.0$\pm$  24.0   &       $<$	  56.0  	    \\ 
          57 &12:18:46.687&+14:21:55.55&3.73&	    $<$  0.7        	          &    0.7$\pm$0.2    &     1.2$\pm$0.2    &		      4.5$\pm$   0.9  & 		  12.7$\pm$   2.0  &	   		    51.6$\pm$	4.8   & 		 66.2$\pm$   6.0   &		     139.0$\pm$   5.0	&		  162.0$\pm$   5.0   &  		84.0$\pm$  11.0   &		    619.0$\pm$  46.0   &		1431.0$\pm$ 416.0   \\ 
          58 &12:18:48.887&+14:22:09.52&4.28&	         5.4   $\pm$	   1.1    &    0.8$\pm$0.2    &     3.0$\pm$0.4    &	  $<$	      1.6	      &        $<$	   3.8  	   &	   		    25.0$\pm$	7.3   & 		 34.6$\pm$   9.8   &		      84.0$\pm$   6.0	&		   66.0$\pm$   4.0   &  		81.0$\pm$  16.0   &	  $<$	     43.0	       &       $<$	 102.0  	    \\ 
          59 &12:18:42.639&+14:21:35.98&3.98&	    $<$  0.5        	          &    0.5$\pm$0.1    &     1.6$\pm$0.1    &		      2.8$\pm$   0.4  & 		   3.3$\pm$   0.6  &	   		     4.4$\pm$	1.1   &       $<$	  1.8		   &		      39.0$\pm$   2.0	&		   40.0$\pm$   3.0   &  		42.0$\pm$   8.0   &	  $<$	     24.0	       &       $<$	  51.0  	    \\ 
          60 &12:18:29.684&+14:22:13.22&5.35&	    $<$  1.0     	          &    0.8$\pm$0.2    &     2.0$\pm$0.2    &		      4.7$\pm$   0.5  & 		  15.2$\pm$   0.7  &	   		    86.2$\pm$	3.2   & 		113.2$\pm$   4.6   &		     161.0$\pm$   4.0	&		  146.0$\pm$   5.0   &  		65.0$\pm$  17.0   &		     88.0$\pm$  21.0   &       $<$	  81.0  	    \\ 
\hline
\end{tabular}}

Notes: $LyC$ is the flux in the Lyman continuum pseudo filter $PSEUDO_{LyC}$ (see Boselli et al. 2016b for details). 
\label{regiondata}
\end{table}
\end{landscape}

\begin{table}
\caption{Input parameters for CIGALE. }
\label{cigale}
{\tiny
\[
\begin{tabular}{ccc}
\hline
\noalign{\smallskip}
\hline
Parameter       &  value                        & Units         \\
\hline
Pop.synth.mod.  & Bruzual \& Charlot (2003)     &               \\
Dust model      & Draine \& Li (2007)           &               \\
IMF             & Salpeter                      &               \\
Metallicity     & 0.02                          &               \\
Age             &  1-301, step 2                & Myr           \\
Extinction law	& Calzetti			&		\\
$E(B-V)$	& 0.0,0.1,0.2,0.3,0.4           & mag           \\
UV bump amplitude& 0, 1.5, 3                    &               \\ 
$Q_{PAH}$       & 0.47, 3.90, 7.32              &               \\
$U_{min}$       & 0.1, 1.0, 5.0, 10.0           &               \\
$\alpha$        & 1.5, 2.0, 2.5                 &               \\
$\gamma$        & 0.01, 0.3, 1.0                &               \\
\noalign{\smallskip}
\hline
\end{tabular}
\]
Notes:\\
Line 1: population synthesis model.\\
Line 2: dust model.\\
Line 3: IMF.\\
Line 4: stellar metallicity.\\
Line 5: age interval and sampling.\\
Line 6: extinction law (Calzetti et al. 2000).\\
Line 7: attenuation.\\
Line 8: amplitude of the UV bump (Noll et al. 2009).\\
Line 9: mass fraction of the dust composed of PAH particles containing $<$ 10$^3$ atoms (Draine \& Li 2007).\\
Line 10: diffuse interstellar radiation field (Draine \& Li 2007).\\
Line 11: power-law index of the interstellar radiation field (Draine \& Li 2007).\\
Line 12: fraction of the dust mass that is close to OB associations and is exposed to a stellar radiation field with intensity $U$ $>$ $U_{min}$ (Draine \& Li 2007).\\
}
\end{table}

\begin{table}
\caption{Parameters derived from the SED fitting. }
\label{cigale_out}
{\tiny
\[
\begin{tabular}{ccccc}
\hline
\noalign{\smallskip}
\hline
Region  & log $M_{star}$ 	& log $SFR$		& age	& Diameter\\
	& M$_{\odot}$ 		& M$_{\odot}$ yr$^{-1}$ & Myr   & pc \\
\hline
           1   &     3.91  $\pm$ 0.02 &     -2.09 $\pm$ 0.03 &       1.02 $\pm$   0.21  &	   -		    \\ 
           2   &     3.94  $\pm$ 0.43 &     -3.23 $\pm$ 0.24 &       8.31 $\pm$   2.94  &	   -		    \\ 
           3   &     4.53  $\pm$ 0.06 &     -3.30 $\pm$ 0.06 &      14.36 $\pm$   0.99  &     483.0$\pm$    4.6     \\ 
           4   &     3.62  $\pm$ 0.36 &     -3.28 $\pm$ 0.14 &       6.07 $\pm$   2.51  &     508.7$\pm$   14.9     \\ 
           5   &     3.62  $\pm$ 0.23 &     -2.79 $\pm$ 0.23 &       3.63 $\pm$   2.86  &     327.1$\pm$    3.2     \\ 
           6   &     3.13  $\pm$ 0.23 &     -3.06 $\pm$ 0.16 &       2.16 $\pm$   2.01  &     129.9$\pm$    3.4     \\ 
           7   &     5.13  $\pm$ 0.14 &     -3.95 $\pm$ 0.14 &      24.74 $\pm$   1.85  &     171.1$\pm$    6.5     \\ 
           8   &     3.62  $\pm$ 0.09 &     -3.49 $\pm$ 0.10 &       8.28 $\pm$   1.52  &     934.1$\pm$   17.6     \\ 
           9   &     3.67  $\pm$ 0.15 &     -3.38 $\pm$ 0.17 &       7.67 $\pm$   2.18  &      75.8$\pm$    4.0     \\ 
          10   &     4.72  $\pm$ 0.25 &     -3.93 $\pm$ 0.17 &      23.18 $\pm$  10.75  &	   -		    \\ 
          11   &     4.81  $\pm$ 0.23 &     -3.80 $\pm$ 0.23 &      22.36 $\pm$   7.44  &     239.6$\pm$    4.2     \\ 
          12   &     4.27  $\pm$ 0.19 &     -3.12 $\pm$ 0.05 &      10.19 $\pm$   2.03  &     295.2$\pm$    2.0     \\ 
          13   &     3.74  $\pm$ 0.10 &     -3.60 $\pm$ 0.06 &      10.22 $\pm$   1.25  &	   -		    \\ 
          14   &     2.70  $\pm$ 0.30 &     -3.69 $\pm$ 0.25 &       3.39 $\pm$   2.50  &	   -		    \\ 
          15   &     3.40  $\pm$ 0.18 &     -3.77 $\pm$ 0.08 &       8.63 $\pm$   1.44  &	   -		    \\ 
          16   &     5.29  $\pm$ 0.42 &     -4.51 $\pm$ 1.45 &      70.54 $\pm$  47.12  &	   -		    \\ 
          17   &     4.99  $\pm$ 0.26 &     -4.43 $\pm$ 0.66 &      34.30 $\pm$  14.07  &	   -		    \\ 
          18   &     4.42  $\pm$ 0.19 &     -4.03 $\pm$ 0.11 &      19.13 $\pm$   2.66  &	   -		    \\ 
          19   &     4.37  $\pm$ 0.16 &     -2.33 $\pm$ 0.27 &       5.42 $\pm$   2.40  &	   -		    \\ 
          20   &     5.82  $\pm$ 0.05 &            -	     &     157.30 $\pm$  38.21  &	   -		    \\ 
          21   &     5.49  $\pm$ 0.03 &     -3.43 $\pm$ 0.07 &      23.36 $\pm$   1.17  &	   -		    \\ 
          22   &     3.93  $\pm$ 0.35 &     -3.66 $\pm$ 0.11 &      11.78 $\pm$   2.82  &	   -		    \\ 
          23   &     4.72  $\pm$ 0.29 &     -3.33 $\pm$ 0.07 &      15.81 $\pm$   2.06  &	   -		    \\ 
          24   &     5.29  $\pm$ 0.34 &     -5.71 $\pm$ 1.78 &     101.90 $\pm$  59.92  &     460.9$\pm$   14.5     \\ 
          25   &     5.02  $\pm$ 0.37 &     -5.29 $\pm$ 1.47 &      77.34 $\pm$  48.48  &     302.6$\pm$    8.3     \\ 
          26   &     4.94  $\pm$ 0.33 &     -5.47 $\pm$ 1.20 &      62.88 $\pm$  35.02  &	   -		    \\ 
          27   &     3.85  $\pm$ 0.60 &     -3.82 $\pm$ 0.14 &      12.33 $\pm$   6.20  &	   -		    \\ 
          28   &     3.61  $\pm$ 0.27 &     -3.09 $\pm$ 0.14 &       4.97 $\pm$   1.55  &	   -		    \\ 
          29   &     5.68  $\pm$ 0.08 &            -	     &     135.04 $\pm$  23.71  &	   -		    \\ 
          30   &     4.81  $\pm$ 0.14 &            -	     &      67.38 $\pm$  19.17  &	   -		    \\ 
          31   &     5.56  $\pm$ 0.11 &     -4.55 $\pm$ 1.11 &      74.04 $\pm$  23.45  &	   -		    \\ 
          32   &     5.58  $\pm$ 0.06 &            -	     &      98.98 $\pm$  14.48  &	   -		    \\ 
          33   &     5.74  $\pm$ 0.05 &            -	     &     189.87 $\pm$  21.04  &	   -		    \\ 
          34   &     6.01  $\pm$ 0.03 &            -	     &     257.35 $\pm$  20.25  &	   -		    \\ 
          35   &     5.19  $\pm$ 0.07 &            -	     &     112.56 $\pm$  39.00  &	   -		    \\ 
          36   &     4.91  $\pm$ 0.08 &     -5.88 $\pm$ 1.10 &      58.65 $\pm$  24.05  &	   -		    \\ 
          37   &     4.37  $\pm$ 0.30 &     -5.02 $\pm$ 0.75 &      36.71 $\pm$  17.70  &	   -		    \\ 
          38   &     3.87  $\pm$ 0.24 &     -4.73 $\pm$ 0.44 &      22.98 $\pm$   7.76  &	   -		    \\ 
          39   &     3.95  $\pm$ 1.31 &     -3.75 $\pm$ 0.35 &      14.90 $\pm$  26.69  &      96.1$\pm$    3.8     \\ 
          40   &     3.59  $\pm$ 0.44 &     -3.95 $\pm$ 0.14 &      11.20 $\pm$   3.23  &	   -		    \\ 
          41   &     3.46  $\pm$ 0.55 &     -3.87 $\pm$ 0.17 &       9.16 $\pm$   3.42  &	   -		    \\ 
          42   &     6.68  $\pm$ 0.02 &            -	     &     295.63 $\pm$  14.78  &	   -		    \\ 
          43   &     4.90  $\pm$ 0.11 &            -	     &      82.68 $\pm$  46.63  &	   -		    \\ 
          44   &     4.92  $\pm$ 0.21 &            -	     &      79.51 $\pm$  31.47  &	   -		    \\ 
          45   &     4.73  $\pm$ 0.32 &     -4.91 $\pm$ 0.81 &      42.22 $\pm$  22.32  &	   -		    \\ 
          46   &     3.80  $\pm$ 0.32 &     -4.26 $\pm$ 0.18 &      16.58 $\pm$   5.32  &	   -		    \\ 
          47   &     4.53  $\pm$ 0.31 &     -6.00 $\pm$ 1.36 &      68.15 $\pm$  37.70  &	   -		    \\ 
          48   &     4.57  $\pm$ 0.35 &     -5.42 $\pm$ 1.10 &      57.26 $\pm$  34.39  &	   -		    \\ 
          49   &     3.78  $\pm$ 0.19 &     -4.85 $\pm$ 0.37 &      22.48 $\pm$   5.93  &	   -		    \\ 
          50   &     5.36  $\pm$ 0.07 &     -5.92 $\pm$ 1.52 &      66.41 $\pm$  20.87  &	   -		    \\ 
          51   &     4.33  $\pm$ 0.18 &     -5.21 $\pm$ 0.75 &      34.70 $\pm$  11.35  &	   -		    \\ 
          52   &     4.18  $\pm$ 0.20 &     -4.88 $\pm$ 0.47 &      27.28 $\pm$   8.08  &	   -		    \\ 
          53   &     2.82  $\pm$ 0.20 &     -3.74 $\pm$ 0.31 &       4.68 $\pm$   2.46  &      77.0$\pm$    7.7     \\ 
          54   &     5.29  $\pm$ 0.09 &            -	     &      84.37 $\pm$  35.80  &	   -		    \\ 
          55   &     4.43  $\pm$ 0.31 &     -5.27 $\pm$ 0.85 &      43.25 $\pm$  22.36  &	   -		    \\ 
          56   &     3.50  $\pm$ 0.07 &     -3.65 $\pm$ 0.08 &       8.65 $\pm$   1.28  &     281.6$\pm$    5.9     \\ 
          57   &     6.35  $\pm$ 0.02 &            -	     &     283.92 $\pm$  15.31  &	   -		    \\ 
          58   &     5.12  $\pm$ 0.51 &     -4.40 $\pm$ 0.41 &      79.20 $\pm$  79.80  &	   -		    \\ 
          59   &     4.90  $\pm$ 0.10 & 	   -	     &      86.99 $\pm$  20.37  &	   -		    \\ 
          60   &     6.40  $\pm$ 0.02 & 	   -	     &     294.89 $\pm$  14.74  &	   -		    \\ 

\noalign{\smallskip}
\hline
\end{tabular}
\]
}
\end{table}


\begin{thebibliography}{}

\bibitem[Ambrocio-Cruz et al.(2016)]{2016MNRAS.457.2048A} Ambrocio-Cruz, P., Le Coarer, E., Rosado, M., et al.\ 2016, \mnras, 457, 2048 
\bibitem[Arrigoni Battaia et al.(2012)]{2012A&A...543A.112A} Arrigoni Battaia, F., Gavazzi, G., Fumagalli, M., et al.\ 2012, \aap, 543, A112 
\bibitem[Behroozi et al.(2010)]{2010ApJ...717..379B} Behroozi, P.~S., Conroy, C., \& Wechsler, R.~H.\ 2010, \apj, 717, 379 
\bibitem[Bekki et al.(2005)]{2005MNRAS.363L..21B} Bekki, K., Koribalski, B.~S., \& Kilborn, V.~A.\ 2005, \mnras, 363, L21 
\bibitem[Bellhouse et al.(2017)]{2017ApJ...844...49B} Bellhouse, C., Jaff{\'e}, Y.~L., Hau, G.~K.~T., et al.\ 2017, \apj, 844, 49 
\bibitem[Bendo et al.(2012)]{2012MNRAS.423..197B} Bendo, G.~J., Galliano, F., \& Madden, S.~C.\ 2012, \mnras, 423, 197 
\bibitem[Binney \& Tremaine(2008)]{2008gady.book.....B} Binney, J., \& Tremaine, S.\ 2008, Galactic Dynamics: Second Edition, by James Binney and Scott Tremaine.~ISBN 978-0-691-13026-2 (HB).~Published by Princeton University Press, Princeton, NJ USA, 2008.,  
\bibitem[Blakeslee et al.(2009)]{2009ApJ...694..556B} Blakeslee, J.~P., Jord{\'a}n, A., Mei, S., et al.\ 2009, \apj, 694, 556 
\bibitem[Blitz \& Rosolowsky(2006)]{2006ApJ...650..933B} Blitz, L., \& Rosolowsky, E.\ 2006, \apj, 650, 933 
\bibitem[Boissier et al.(2012)]{2012A&A...545A.142B} Boissier, S., Boselli, A., Duc, P.-A., et al.\ 2012, \aap, 545, A142 
\bibitem[Boquien et al.(2007)]{2007A&A...467...93B} Boquien, M., Duc, P.-A., Braine, J., et al.\ 2007, \aap, 467, 93 
\bibitem[Boquien et al.(2009)]{2009AJ....137.4561B} Boquien, M., Duc, P.-A., Wu, Y., et al.\ 2009, \aj, 137, 4561 
\bibitem[Boquien et al.(2010)]{2010AJ....140.2124B} Boquien, M., Duc, P.-A., Galliano, F., et al.\ 2010, \aj, 140, 2124 
\bibitem[Boquien et al.(2011)]{2011A&A...533A..19B} Boquien, M., Lisenfeld, U., Duc, P.-A., et al.\ 2011, \aap, 533, A19 
\bibitem[Boquien et al.(2012)]{2012A&A...539A.145B} Boquien, M., Buat, V., Boselli, A., et al.\ 2012, \aap, 539, A145 
\bibitem[Boquien et al.(2014)]{2014A&A...571A..72B} Boquien, M., Buat, V., \& Perret, V.\ 2014, \aap, 571, A72 
\bibitem{2006PASP..118..517B} Boselli, A., \& Gavazzi, G.\ 2006, \pasp, 118, 517 
\bibitem[Boselli \& Gavazzi(2014)]{2014A&ARv..22...74B} Boselli, A., \& Gavazzi, G.\ 2014, \aapr, 22, 74 
\bibitem{2006ApJ...651..811B} Boselli, A., Boissier, S., Cortese, L., et al.\ 2006, \apj, 651, 811 
\bibitem{2008ApJ...674..742B} Boselli, A., Boissier, S., Cortese, L., \& Gavazzi, G.\ 2008a, \apj, 674, 742 
\bibitem{2008A&A...489.1015B} Boselli, A., Boissier, S., Cortese, L., \& Gavazzi, G.\ 2008b, \aap, 489, 1015 
\bibitem[Boselli et al.(2009)]{2009ApJ...706.1527B} Boselli, A., Boissier, S., Cortese, L., et al.\ 2009, \apj, 706, 1527 
\bibitem[Boselli et al.(2010)]{2010PASP..122..261B} Boselli, A., Eales, S., Cortese, L., et al.\ 2010, \pasp, 122, 261 
\bibitem{2011A&A...528A.107B} Boselli, A., Boissier, S., Heinis, S., et al.\ 2011, \aap, 528, A107 
\bibitem[Boselli et al.(2014)]{2014A&A...564A..67B} Boselli, A., Cortese, L., Boquien, M., et al.\ 2014a, \aap, 564, A67 
\bibitem[Boselli et al.(2014)]{2014A&A...570A..69B} Boselli, A., Voyer, E., Boissier, S., et al.\ 2014b, \aap, 570, AA69 
\bibitem[Boselli et al.(2016)]{2016A&A...587A..68B} Boselli, A., Cuillandre, J.~C., Fossati, M., et al.\ 2016a, \aap, 587, A68 
\bibitem[Boselli et al.(2016)]{2016A&A...596A..11B} Boselli, A., Roehlly, Y., Fossati, M., et al.\ 2016b, \aap, 596, A11 
\bibitem[Boselli et al.(2018)]{} Boselli, A., Fossati, M., Ferrarese, L., et al., 2018, A\&A, in press (https://arxiv.org/abs/1802.02829) (paper I)
\bibitem[Bresolin(2017)]{2017ASSL..434..145B} Bresolin, F.\ 2017, Outskirts of Galaxies, 434, 145 
\bibitem[Bresolin et al.(2009)]{2009ApJ...695..580B} Bresolin, F., Ryan-Weber, E., Kennicutt, R.~C., \& Goddard, Q.\ 2009, \apj, 695, 580 
\bibitem[Bresolin et al.(2012)]{2012ApJ...750..122B} Bresolin, F., Kennicutt, R.~C., \& Ryan-Weber, E.\ 2012, \apj, 750, 122 
\bibitem[Bruzual \& Charlot(2003)]{2003MNRAS.344.1000B} Bruzual, G., \& Charlot, S.\ 2003, \mnras, 344, 1000 
\bibitem[Calzetti et al.(2000)]{2000ApJ...533..682C} Calzetti, D., Armus, L., Bohlin, R.~C., et al.\ 2000, \apj, 533, 682 
\bibitem[Cayatte et al.(1990)]{1990AJ....100..604C} Cayatte, V., van Gorkom, J.~H., Balkowski, C., \& Kotanyi, C.\ 1990, \aj, 100, 604 
\bibitem[Cayatte et al.(1994)]{1994AJ....107.1003C} Cayatte, V., Kotanyi, C., Balkowski, C., \& van Gorkom, J.~H.\ 1994, \aj, 107, 1003 
\bibitem[Cervi{\~n}o \& Luridiana(2004)]{2004A&A...413..145C} Cervi{\~n}o, M., \& Luridiana, V.\ 2004, \aap, 413, 145 
\bibitem[Cervi{\~n}o et al.(2003)]{2003A&A...407..177C} Cervi{\~n}o, M., Luridiana, V., P{\'e}rez, E., V{\'{\i}}lchez, J.~M., \& Valls-Gabaud, D.\ 2003, \aap, 407, 177 
\bibitem[Cervi{\~n}o et al.(2013)]{2013A&A...553A..31C} Cervi{\~n}o, M., Rom{\'a}n-Z{\'u}{\~n}iga, C., Luridiana, V., et al.\ 2013, \aap, 553, A31 
\bibitem[Chung et al.(2007)]{2007ApJ...659L.115C} Chung, A., van Gorkom, J.~H., Kenney, J.~D.~P., \& Vollmer, B.\ 2007, \apjl, 659, L115 
\bibitem[Chung et al.(2009)]{2009AJ....138.1741C} Chung, A., van Gorkom, J.~H., Kenney, J.~D.~P., Crowl, H., \& Vollmer, B.\ 2009, \aj, 138, 1741 
\bibitem[Ciesla et al.(2012)]{2012A&A...543A.161C} Ciesla, L., Boselli, A., Smith, M.~W.~L., et al.\ 2012, \aap, 543, A161 
\bibitem[Ciesla et al.(2014)]{2014A&A...565A.128C} Ciesla, L., Boquien, M., Boselli, A., et al.\ 2014, \aap, 565, A128 
\bibitem[Consolandi et al. 2017]{} Consolandi G., Gavazzi G., Fossati M., et al.: arXiv:1707.06241
\bibitem[Copetti et al.(1985)]{1985A&A...152..427C} Copetti, M.~V.~F., Pastoriza, M.~G., \& Dottori, H.~A.\ 1985, \aap, 152, 427 
\bibitem[Cortese et al.(2006)]{2006A&A...453..847C} Cortese, L., Gavazzi, G., Boselli, A., et al.\ 2006, \aap, 453, 847 
\bibitem[Cortese et al.(2010)]{2010A&A...518L..49C} Cortese, L., Davies, J.~I., Pohlen, M., et al.\ 2010, \aap, 518, L49 
\bibitem[Cortese et al.(2012a)]{2012A&A...540A..52C} Cortese, L., Ciesla, L., Boselli, A., et al.\ 2012a, \aap, 540, A52 
\bibitem[Cortese et al.(2012b)]{2012A&A...544A.101C} Cortese, L., Boissier, S., Boselli, A., et al.\ 2012b, \aap, 544, A101 
\bibitem[Cortese et al.(2014)]{2014MNRAS.440..942C} Cortese, L., Fritz, J., Bianchi, S., et al.\ 2014, \mnras, 440, 942 
\bibitem[Cowie \& Songaila(1977)]{1977Natur.266..501C} Cowie, L.~L., \& Songaila, A.\ 1977, \nat, 266, 501 
\bibitem[da Silva et al.(2012)]{2012ApJ...745..145D} da Silva, R.~L., Fumagalli, M., \& Krumholz, M.\ 2012, \apj, 745, 145 
\bibitem[Davies et al.(2004)]{2004MNRAS.349..922D} Davies, J., Minchin, R., Sabatini, S., et al.\ 2004, \mnras, 349, 922 
\bibitem[Davies et al.(2010)]{2010A&A...518L..48D} Davies, J.~I., Baes, M., Bendo, G.~J., et al.\ 2010, \aap, 518, L48 
\bibitem[Draine \& Li(2007)]{2007ApJ...657..810D} Draine, B.~T., \& Li, A.\ 2007, \apj, 657, 810 
\bibitem[Dressler(1980)]{1980ApJ...236..351D} Dressler, A.\ 1980, \apj, 236, 351 
\bibitem[Dressler(2004)]{2004cgpc.symp..206D} Dressler, A.\ 2004, Clusters of Galaxies: Probes of Cosmological Structure and Galaxy Evolution, 206 
\bibitem[Duc \& Bournaud(2008)]{2008ApJ...673..787D} Duc, P.-A., \& Bournaud, F.\ 2008, \apj, 673, 787-797 
\bibitem[Elmegreen(2000)]{2000ApJ...539..342E} Elmegreen, B.~G.\ 2000, \apj, 539, 342 
\bibitem[Engargiola et al.(2003)]{2003ApJS..149..343E} Engargiola, G., Plambeck, R.~L., Rosolowsky, E., \& Blitz, L.\ 2003, \apjs, 149, 343 
\bibitem[Ferrarese et al.(2012)]{2012ApJS..200....4F} Ferrarese, L., C{\^o}t{\'e}, P., Cuillandre, J.-C., et al.\ 2012, \apjs, 200, 4 
\bibitem[Fitzpatrick(1999)]{1999PASP..111...63F} Fitzpatrick, E.~L.\ 1999, \pasp, 111, 63 
\bibitem[Fossati et al.(2012)]{2012A&A...544A.128F} Fossati, M., Gavazzi, G., Boselli, A., \& Fumagalli, M.\ 2012, \aap, 544, A128 
\bibitem[Fossati et al.(2016)]{2016MNRAS.455.2028F} Fossati, M., Fumagalli, M., Boselli, A., et al.\ 2016, \mnras, 455, 2028 
\bibitem[Fossati et al.(2018)]{} Fossati, M., Mendel, J.T., Boselli, A., et al., A\&A, in press (https://arxiv.org/abs/1801.09685) (paper II)
\bibitem[Fritz et al.(2017)]{2017ApJ...848..132F} Fritz, J., Moretti, A., Gullieuszik, M., et al.\ 2017, \apj, 848, 132 
\bibitem[Fumagalli et al.(2009)]{2009ApJ...697.1811F} Fumagalli, M., Krumholz, M.~R., Prochaska, J.~X., Gavazzi, G., \& Boselli, A.\ 2009, \apj, 697, 1811 
\bibitem[Fumagalli et al.(2011)]{2011A&A...528A..46F} Fumagalli, M., Gavazzi, G., Scaramella, R., \& Franzetti, P.\ 2011a, \aap, 528, A46 
\bibitem[Fumagalli et al.(2011)]{2011ApJ...741L..26F} Fumagalli, M., da Silva, R.~L., \& Krumholz, M.~R.\ 2011b, \apjl, 741, L26 
\bibitem[Fumagalli et al.(2014)]{2014MNRAS.445.4335F} Fumagalli, M., Fossati, M., Hau, G.~K.~T., et al.\ 2014, \mnras, 445, 4335 
\bibitem[Gavazzi \& Jaffe(1987)]{1987A&A...186L...1G} Gavazzi, G., \& Jaffe, W.\ 1987, \aap, 186, L1 
\bibitem[Gavazzi \& Jaffe(1985)]{1985ApJ...294L..89G} Gavazzi, G., \& Jaffe, W.\ 1985, \apjl, 294, L89 
\bibitem[Gavazzi et al.(1995)]{1995A&A...304..325G} Gavazzi, G., Contursi, A., Carrasco, L., et al.\ 1995, \aap, 304, 325 
\bibitem[Gavazzi et al.(1998)]{1998AJ....115.1745G} Gavazzi, G., Catinella, B., Carrasco, L., Boselli, A., \& Contursi, A.\ 1998, \aj, 115, 1745 
\bibitem[Gavazzi et al.(1999)]{1999MNRAS.304..595G} Gavazzi, G., Boselli, A., Scodeggio, M., Pierini, D., \& Belsole, E.\ 1999, \mnras, 304, 595 
\bibitem[Gavazzi et al.(2001)]{2001ApJ...563L..23G} Gavazzi, G., Boselli, A., Mayer, L., et al.\ 2001, \apjl, 563, L23 
\bibitem[Gavazzi et al.(2003)]{2003A&A...400..451G} Gavazzi, G., Boselli, A., Donati, A., Franzetti, P., \& Scodeggio, M.\ 2003, \aap, 400, 451 
\bibitem[Gavazzi et al.(2005)]{2005A&A...429..439G} Gavazzi, G., Boselli, A., van Driel, W., \& O'Neil, K.\ 2005, \aap, 429, 439 
\bibitem[Gavazzi et al.(2010)]{2010A&A...517A..73G} Gavazzi, G., Fumagalli, M., Cucciati, O., \& Boselli, A.\ 2010, \aap, 517, A73 
\bibitem[Gavazzi et al.(2013)]{2013A&A...553A..89G} Gavazzi, G., Fumagalli, M., Fossati, M., et al.\ 2013, \aap, 553, A89 
\bibitem[Giovanelli et al.(2005)]{2005AJ....130.2598G} Giovanelli, R., Haynes, M.~P., Kent, B.~R., et al.\ 2005, \aj, 130, 2598 
\bibitem[Giovannoli et al.(2011)]{2011A&A...525A.150G} Giovannoli, E., Buat, V., Noll, S., Burgarella, D., \& Magnelli, B.\ 2011, \aap, 525, A150 
\bibitem[G{\'o}mez et al.(2003)]{2003ApJ...584..210G} G{\'o}mez, P.~L., Nichol, R.~C., Miller, C.~J., et al.\ 2003, \apj, 584, 210 
\bibitem[Gunn \& Gott(1972)]{1972ApJ...176....1G} Gunn, J.~E., \& Gott, J.~R., III 1972, \apj, 176, 1 
\bibitem[Gwyn(2008)]{2008PASP..120..212G} Gwyn, S.~D.~J.\ 2008, \pasp, 120, 212 
\bibitem[Haynes \& Giovanelli(1984)]{1984AJ.....89..758H} Haynes, M.~P., \& Giovanelli, R.\ 1984, \aj, 89, 758 
\bibitem[Haynes et al.(2007)]{2007ApJ...665L..19H} Haynes, M.~P., Giovanelli, R., \& Kent, B.~R.\ 2007, \apjl, 665, L19 
\bibitem[Hester et al.(2010)]{2010ApJ...716L..14H} Hester, J.~A., Seibert, M., Neill, J.~D., et al.\ 2010, \apjl, 716, L14 
\bibitem[Ho et al.(1997)]{1997ApJ...487..579H} Ho, L.~C., Filippenko, A.~V., \& Sargent, W.~L.~W.\ 1997, \apj, 487, 579 
\bibitem[Hughes et al.(2013)]{2013A&A...550A.115H} Hughes, T.~M., Cortese, L., Boselli, A., Gavazzi, G., \& Davies, J.~I.\ 2013, \aap, 550, A115 
\bibitem[J{\'a}chym et al.(2013)]{2013A&A...556A..99J} J{\'a}chym, P., Kenney, J.~D.~P., R{\v z}ui{\v c}ka, A., et al.\ 2013, \aap, 556, A99 
\bibitem[J{\'a}chym et al.(2014)]{2014ApJ...792...11J} J{\'a}chym, P., Combes, F., Cortese, L., Sun, M., \& Kenney, J.~D.~P.\ 2014, \apj, 792, 11 
\bibitem[J{\'a}chym et al.(2017)]{2017ApJ...839..114J} J{\'a}chym, P., Sun, M., Kenney, J.~D.~P., et al.\ 2017, \apj, 839, 114 
\bibitem[Kantharia et al.(2008)]{2008MNRAS.383..173K} Kantharia, N.~G., Rao, A.~P., \& Sirothia, S.~K.\ 2008, \mnras, 383, 173 
\bibitem[Kapferer et al.(2009)]{2009A&A...499...87K} Kapferer, W., Sluka, C., Schindler, S., Ferrari, C., \& Ziegler, B.\ 2009, \aap, 499, 87 
\bibitem[Kenney et al.(2004)]{2004AJ....127.3361K} Kenney, J.~D.~P., van Gorkom, J.~H., \& Vollmer, B.\ 2004, \aj, 127, 3361 
\bibitem[Kenney et al.(2008)]{2008ApJ...687L..69K} Kenney, J.~D.~P., Tal, T., Crowl, H.~H., Feldmeier, J., \& Jacoby, G.~H.\ 2008, \apjl, 687, L69 
\bibitem[Kenney et al.(2014)]{2014ApJ...780..119K} Kenney, J.~D.~P., Geha, M., J{\'a}chym, P., et al.\ 2014, \apj, 780, 119 
\bibitem[Kennicutt(1983)]{1983AJ.....88..483K} Kennicutt, R.~C., Jr.\ 1983, \aj, 88, 483 
\bibitem[Kennicutt et al.(1989)]{1989AJ.....97.1022K} Kennicutt, R.~C., Jr., Keel, W.~C., \& Blaha, C.~A.\ 1989, \aj, 97, 1022 
\bibitem[Kennicutt et al.(2003)]{2003PASP..115..928K} Kennicutt, R.~C., Jr., Armus, L., Bendo, G., et al.\ 2003, \pasp, 115, 928 
\bibitem[Kewley et al.(2001)]{2001ApJ...556..121K} Kewley, L.~J., Dopita, M.~A., Sutherland, R.~S., Heisler, C.~A., \& Trevena, J.\ 2001, \apj, 556, 121 
\bibitem[Koda et al.(2012)]{2012ApJ...749...20K} Koda, J., Yagi, M., Boissier, S., et al.\ 2012, \apj, 749, 20 
\bibitem[Larson et al.(1980)]{1980ApJ...237..692L} Larson, R.~B., Tinsley, B.~M., \& Caldwell, C.~N.\ 1980, \apj, 237, 692 
\bibitem[Lee et al.(2011)]{2011ApJ...735...75L} Lee, J.~H., Hwang, N., \& Lee, M.~G.\ 2011, \apj, 735, 75 
\bibitem[Leroy et al.(2009)]{2009AJ....137.4670L} Leroy, A.~K., Walter, F., Bigiel, F., et al.\ 2009, \aj, 137, 4670 
\bibitem[Lisenfeld et al.(2016)]{2016A&A...590A..92L} Lisenfeld, U., Braine, J., Duc, P.~A., et al.\ 2016, \aap, 590, A92 
\bibitem[Liu et al.(2015)]{2015ApJ...812...34L} Liu, C., Peng, E.~W., C{\^o}t{\'e}, P., et al.\ 2015, \apj, 812, 34 
\bibitem[Liu et al.(2016)]{2016ApJ...830...99L} Liu, Y., Peng, E.~W., Lim, S., et al.\ 2016, \apj, 830, 99 
\bibitem[McCall et al.(1985)]{1985ApJS...57....1M} McCall, M.~L., Rybski, P.~M., \& Shields, G.~A.\ 1985, \apjs, 57, 1 
\bibitem[McLaughlin(1999)]{1999ApJ...512L...9M} McLaughlin, D.~E.\ 1999, \apjl, 512, L9 
\bibitem[Mei et al.(2007)]{2007ApJ...655..144M} Mei, S., Blakeslee, J.~P., C{\^o}t{\'e}, P., et al.\ 2007, \apj, 655, 144 
\bibitem[Mihos et al.(2005)]{2005ApJ...631L..41M} Mihos, J.~C., Harding, P., Feldmeier, J., \& Morrison, H.\ 2005, \apjl, 631, L41 
\bibitem[Mihos et al.(2009)]{2009ApJ...698.1879M} Mihos, J.~C., Janowiecki, S., Feldmeier, J.~J., Harding, P., \& Morrison, H.\ 2009, \apj, 698, 1879 
\bibitem[Mihos et al.(2017)]{2017ApJ...834...16M} Mihos, J.~C., Harding, P., Feldmeier, J.~J., et al.\ 2017, \apj, 834, 16 
\bibitem[Minchin et al.(2005)]{2005ApJ...622L..21M} Minchin, R., Davies, J., Disney, M., et al.\ 2005, \apjl, 622, L21 
\bibitem[Moore et al.(1998)]{1998ApJ...495..139M} Moore, B., Lake, G., \& Katz, N.\ 1998, \apj, 495, 139 
\bibitem[Morrissey et al.(2007)]{2007ApJS..173..682M} Morrissey, P., Conrow, T., Barlow, T.~A., et al.\ 2007, \apjs, 173, 682 
\bibitem[Noll et al.(2009)]{2009A&A...507.1793N} Noll, S., Burgarella, D., Giovannoli, E., et al.\ 2009, \aap, 507, 1793 
\bibitem[Peng et al.(2002)]{2002AJ....124..266P} Peng, C.~Y., Ho, L.~C., Impey, C.~D., \& Rix, H.-W.\ 2002, \aj, 124, 266 
\bibitem[Peng et al.(2010)]{2010AJ....139.2097P} Peng, C.~Y., Ho, L.~C., Impey, C.~D., \& Rix, H.-W.\ 2010, \aj, 139, 2097 
\bibitem[Peng et al.(2006)]{2006ApJ...639..838P} Peng, E.~W., C{\^o}t{\'e}, P., Jord{\'a}n, A., et al.\ 2006, \apj, 639, 838 
\bibitem[Pettini \& Pagel(2004)]{2004MNRAS.348L..59P} Pettini, M., \& Pagel, B.~E.~J.\ 2004, \mnras, 348, L59 
\bibitem[Phookun et al.(1993)]{1993ApJ...418..113P} Phookun, B., Vogel, S.~N., \& Mundy, L.~G.\ 1993, \apj, 418, 113 
\bibitem[Poggianti et al.(2017)]{2017ApJ...844...48P} Poggianti, B.~M., Moretti, A., Gullieuszik, M., et al.\ 2017, \apj, 844, 48 
\bibitem[Roediger \& Br{\"u}ggen(2006)]{2006MNRAS.369..567R} Roediger, E., \& Br{\"u}ggen, M.\ 2006, \mnras, 369, 567 
\bibitem[Roediger \& Br{\"u}ggen(2007)]{2007MNRAS.380.1399R} Roediger, E., \& Br{\"u}ggen, M.\ 2007, \mnras, 380, 1399 
\bibitem[Roediger \& Br{\"u}ggen(2008)]{2008MNRAS.388L..89R} Roediger, E., \& Br{\"u}ggen, M.\ 2008, \mnras, 388, L89 
\bibitem[Roediger \& Hensler(2005)]{2005A&A...433..875R} Roediger, E., \& Hensler, G.\ 2005, \aap, 433, 875 
\bibitem[S{\'a}nchez et al.(2015)]{2015A&A...574A..47S} S{\'a}nchez, S.~F., P{\'e}rez, E., Rosales-Ortega, F.~F., et al.\ 2015, \aap, 574, A47 
\bibitem[S{\'a}nchez-Menguiano et al.(2016)]{2016A&A...587A..70S} S{\'a}nchez-Menguiano, L., S{\'a}nchez, S.~F., P{\'e}rez, I., et al.\ 2016, \aap, 587, A70 
\bibitem[Sand et al.(2017)]{2017ApJ...843..134S} Sand, D.~J., Seth, A.~C., Crnojevi{\'c}, D., et al.\ 2017, \apj, 843, 134 
\bibitem[Sawicki(2012)]{2012PASP..124.1208S} Sawicki, M.\ 2012, \pasp, 124, 1208 
\bibitem[Schlafly \& Finkbeiner(2011)]{2011ApJ...737..103S} Schlafly, E.~F., \& Finkbeiner, D.~P.\ 2011, \apj, 737, 103 
\bibitem[Scott et al.(2012)]{2012MNRAS.419L..19S} Scott, T.~C., Cortese, L., Brinks, E., et al.\ 2012, \mnras, 419, L19 
\bibitem[Sivanandam et al.(2014)]{2014ApJ...796...89S} Sivanandam, S., Rieke, M.~J., \& Rieke, G.~H.\ 2014, \apj, 796, 89 
\bibitem[Skillman et al.(1996)]{1996ApJ...462..147S} Skillman, E.~D., Kennicutt, R.~C., Jr., Shields, G.~A., \& Zaritsky, D.\ 1996, \apj, 462, 147 
\bibitem[Solanes et al.(2001)]{2001ApJ...548...97S} Solanes, J.~M., Manrique, A., Garc{\'{\i}}a-G{\'o}mez, C., et al.\ 2001, \apj, 548, 97 
\bibitem[Solomon et al.(1987)]{1987ApJ...319..730S} Solomon, P.~M., Rivolo, A.~R., Barrett, J., \& Yahil, A.\ 1987, \apj, 319, 730 
\bibitem[Sun et al.(2006)]{2006ApJ...637L..81S} Sun, M., Jones, C., Forman, W., et al.\ 2006, \apjl, 637, L81 
\bibitem[Sun et al.(2007)]{2007ApJ...671..190S} Sun, M., Donahue, M., \& Voit, G.~M.\ 2007, \apj, 671, 190 
\bibitem[Sun et al.(2010)]{2010ApJ...708..946S} Sun, M., Donahue, M., Roediger, E., et al.\ 2010, \apj, 708, 946 
\bibitem[Thilker et al.(2007)]{2007ApJS..173..538T} Thilker, D.~A., Bianchi, L., Meurer, G., et al.\ 2007, \apjs, 173, 538 
\bibitem[Tonnesen \& Bryan(2009)]{2009ApJ...694..789T} Tonnesen, S., \& Bryan, G.~L.\ 2009, \apj, 694, 789 
\bibitem[Tonnesen \& Bryan(2010)]{2010ApJ...709.1203T} Tonnesen, S., \& Bryan, G.~L.\ 2010, \apj, 709, 1203 
\bibitem[Tonnesen \& Bryan(2012)]{2012MNRAS.422.1609T} Tonnesen, S., \& Bryan, G.~L.\ 2012, \mnras, 422, 1609 
\bibitem[Tremblin et al.(2014)]{2014A&A...568A...4T} Tremblin, P., Anderson, L.~D., Didelon, P., et al.\ 2014, \aap, 568, A4 
\bibitem[Verdugo et al.(2015)]{2015A&A...582A...6V} Verdugo, C., Combes, F., Dasyra, K., Salom{\'e}, P., \& Braine, J.\ 2015, \aap, 582, A6 
\bibitem[Vollmer et al.(2005)]{2005A&A...439..921V} Vollmer, B., Huchtmeier, W., \& van Driel, W.\ 2005, \aap, 439, 921 
\bibitem[Vollmer et al.(2006)]{2006A&A...453..883V} Vollmer, B., Soida, M., Otmianowska-Mazur, K., et al.\ 2006, \aap, 453, 883 
\bibitem[Vollmer et al.(2008)]{2008A&A...483...89V} Vollmer, B., Soida, M., Chung, A., et al.\ 2008a, \aap, 483, 89 
\bibitem[Vollmer et al.(2008)]{2008A&A...491..455V} Vollmer, B., Braine, J., Pappalardo, C., \& Hily-Blant, P.\ 2008b, \aap, 491, 455 
\bibitem[Vollmer et al.(2009)]{2009A&A...496..669V} Vollmer, B., Soida, M., Chung, A., et al.\ 2009, \aap, 496, 669 
\bibitem[Vollmer et al.(2012)]{2012A&A...537A.143V} Vollmer, B., Soida, M., Braine, J., et al.\ 2012, \aap, 537, A143 
\bibitem[Voyer et al.(2014)]{2014A&A...569A.124V} Voyer, E.~N., Boselli, A., Boissier, S., et al.\ 2014, \aap, 569, A124 
\bibitem[Xu et al.(2005)]{2005ApJ...619L..11X} Xu, C.~K., Donas, J., Arnouts, S., et al.\ 2005, \apjl, 619, L11 
\bibitem[Yagi et al.(2007)]{2007ApJ...660.1209Y} Yagi, M., Komiyama, Y., Yoshida, M., et al.\ 2007, \apj, 660, 1209 
\bibitem[Yagi et al.(2010)]{2010AJ....140.1814Y} Yagi, M., Yoshida, M., Komiyama, Y., et al.\ 2010, \aj, 140, 1814 
\bibitem[Yagi et al.(2013)]{2013ApJ...778...91Y} Yagi, M., Gu, L., Fujita, Y., et al.\ 2013, \apj, 778, 91 
\bibitem[Yagi et al.(2017)]{2017ApJ...839...65Y} Yagi, M., Yoshida, M., Gavazzi, G., et al.\ 2017, \apj, 839, 65 
\bibitem[Yoshida et al.(2002)]{2002ApJ...567..118Y} Yoshida, M., Yagi, M., Okamura, S., et al.\ 2002, \apj, 567, 118 
\bibitem[Yoshida et al.(2012)]{2012ApJ...749...43Y} Yoshida, M., Yagi, M., Komiyama, Y., et al.\ 2012, \apj, 749, 43 
\bibitem[Whitmore et al.(1993)]{1993ApJ...407..489W} Whitmore, B.~C., Gilmore, D.~M., \& Jones, C.\ 1993, \apj, 407, 489 
\bibitem[Wilson et al.(2011)]{2011MNRAS.410.1409W} Wilson, C.~D., Warren, B.~E., Irwin, J., et al.\ 2011, \mnras, 410, 1409 
\bibitem[Wong et al.(2006)]{2006MNRAS.371.1855W} Wong, O.~I., Ryan-Weber, E.~V., Garcia-Appadoo, D.~A., et al.\ 2006, \mnras, 371, 1855 
\bibitem[Wright et al.(2010)]{2010AJ....140.1868W} Wright, E.~L., Eisenhardt, P.~R.~M., Mainzer, A.~K., et al.\ 2010, \aj, 140, 1868-1881 
\bibitem[Zhang et al.(2013)]{2013ApJ...777..122Z} Zhang, B., Sun, M., Ji, L., et al.\ 2013, \apj, 777, 122 



\end{thebibliography}
\end{document}